\documentclass[11pt,a4paper]{article}
\usepackage[utf8]{inputenc}
\usepackage[T1]{fontenc}
\usepackage{geometry}
\usepackage{url}            % simple URL typesetting
\usepackage{booktabs}       % professional-quality tables
\usepackage{amsfonts}       % blackboard math symbols
\usepackage{nicefrac}       % compact symbols for 1/2, etc.
\usepackage{microtype}      % microtypography
\usepackage{lipsum}
\usepackage{fancyhdr}       % header

\usepackage{amsmath,amssymb,amsbsy}
\usepackage{graphicx}    % 插入图片
\usepackage{subcaption}  % 子图排版
\usepackage{enumitem}    % 列表优化
\usepackage{float}       % 浮动体固定
\usepackage{multirow}    % 多行列
\usepackage{xcolor}      % 颜色
\usepackage{listings}    % 代码块
\usepackage{appendix}    % 附录
\usepackage{hyperref}    % 超链接（arXiv推荐）
\usepackage{natbib}      % 参考文献（numbers格式）
\usepackage{authblk} % 用于作者和单位设置
\usepackage{algorithm}
\usepackage{algorithmic}
\hypersetup{colorlinks=true, linkcolor=blue, citecolor=blue, urlcolor=blue}
\usepackage{authblk}
\usepackage{amsmath}

\begin{document}

% 论文标题
% \title{MPI-Q: A Message Communication Library for Large-Scale Classical-Quantum Heterogeneous Hybrid Distributed Computing}
\title{\textbf{MPI-Q: A Message Communication Library for Large-Scale Classical-Quantum Heterogeneous Hybrid Distributed Computing}}

% --- 作者列表 ---
\author{
  Feng Wang$^{1,}$\thanks{These authors contributed equally to this work.},
  Junchao Wang$^{1,}$\thanks{These authors contributed equally to this work.}\textsuperscript{ ,}\thanks{Corresponding author: wangjunchao11@126.com},
  Zeyuan Wang$^{1}$,
  Lei Li$^{1}$,
  Hang Lian$^{1}$,
  Yangyang Fei$^{1}$,
  Jinyang Yao$^{1}$,
  Xuyan Qi$^{1}$,
  Fudong Liu$^{1}$,
  Yifan Hou$^{1}$,
  Shibo Liang$^{1}$,
  Zheng Shan$^{1,}$\thanks{Corresponding author: shanzhengzz@163.com}
}
\affil{
  $^{1}$Laboratory for Advanced Computing and Intelligence Engineering,
  Information Engineering University,\\
  Zhengzhou, CN
}

\date{\today} % 可替换为提交日期，如\date{March 2026}
\maketitle

% 摘要
\begin{abstract}
The classical-quantum system heterogeneity (different data characteristics, execution paradigms and synchronization mechanism etc.) renders existing distributed communication mechanisms (e.g. MPI, NCCL etc.) inadequate. This bottleneck severely impairs operational synergy and programming efficiency. Thus, the performance of hybrid applications on classical-quantum heterogeneous infrastructures is directly limited.

To address these challenges, this paper proposes a message-passing library tailored for large-scale classical-quantum heterogeneous distributed computing, referred to as MPI-Q. The design centers on three mechanisms. First, it defines a heterogeneous hybrid communication domain that achieves unified management of classical and quantum processes in heterogeneous hybrid systems. Second, it uses a lightweight communication path that allows classical control nodes to send device-ready waveform data directly to quantum MonitorProcesses, avoiding unnecessary relay stages. Third, it establishes a heterogeneous hybrid synchronization mechanism to tackle the problem of timing control for multi-node quantum operations. While retaining the traditional MPI programming model, MPI-Q achieves extension toward quantum subsystems. Experiments on distributed GHZ state preparation demonstrate that this model exhibits near-linear scalability, achieving a maximum speedup of 18.76× on 24 quantum nodes. This proves that the library can effectively support large-scale heterogeneous hybrid distributed computing applications, filling the technical gap in this field.

\end{abstract}

% 关键词
\textbf{Keywords:} Classical-quantum heterogeneous hybrid; Distributed computing; Message passing interface library; Large-scale distributed computing; Classical-quantum heterogenous hybrid integration; Shared control architectures; Quantum control system; Heterogeneous programmability

\vspace{0.5cm} % 空行分隔

% 正文开始
\section{Introduction}
Quantum computing, with its unique parallel computing capability, is expected to solve complex problems that are intractable for classical computers~\cite{kumar2025secure, cicero_simulation_2025, singh_role_2023}, such as large integer factorization~\cite{shor1994algorithms} and quantum chemistry simulation~\cite{herrmann2024quantum}.
However, the realization of practical quantum applications typically requires the support of millions of physical qubits~\cite{sivak2022quantum, mohseni2024how}. Currently, developing a monolithic quantum processor with millions of qubits remains highly challenging. This is severely limited by various physical constraints. Specifically, these constraints include dilution refrigerator space~\cite{sivak2022quantum, mohseni2024how}, fabrication yield~\cite{gavande2024tackling, saraiva2022materials}, qubit coherence times~\cite{bernien2020probing, hao2025optical}, gate fidelities~\cite{dong2021fast, coelloperez2022quantum, motzoi2009simple}, and integration scale limits~\cite{grumbling2019quantum, ding2020systematic, smith2022scaling}. Therefore, a feasible solution is to construct a classical-quantum heterogeneous hybrid system to achieve classical-quantum integration~\cite{zappin2025quantum, bensoussan2025taxonomy, mcclean2016theory, esposito2025slurm}. This approach integrates multiple small-scale quantum nodes~\cite{chia2024hybrid, gazda2024pragma} via coherent interconnections~\cite{esposito2025slurm, chia2024hybrid, pasini2025enabling}. Consequently, this architecture not only significantly expands the total number of accessible qubits~\cite{chia2024hybrid}. Furthermore, it reduces the overall task execution time through parallel computing~\cite{bravomontes2025architectural}.

This trend makes classical-quantum heterogeneous computing increasingly important. The essence of such synergistic computing is a heterogeneous hybrid architecture. This architecture consists of a distributed classical computing cluster and a distributed quantum computing system. In
such systems, classical clusters orchestrate control flow, data movement, and iterative optimization, while quantum processors execute selected kernels. The challenge is that existing distributed communication models do not match this execution structure. This is because a fundamental difference exists between classical and quantum data. Classical data is characterized by determinism and replicability. In stark contrast, quantum data relies on the superposition and entanglement states of qubits~\cite{mermin1990extreme}. It exhibits probabilistic behavior, adheres to the no-cloning theorem, and is highly susceptible to environmental noise~\cite{zurek2003decoherence, miyadera2009nocloning}. As a result, a classical-quantum system cannot be managed as a
conventional data-centric distributed platform~\cite{saxena2021distributed, seiwerth2025extending}

To solve this problem, this paper proposes a novel distributed classical-quantum heterogeneous hybrid architecture. The core feature of this architecture is shared control. Specifically, the architecture consists of multiple distributed computing nodes. Furthermore, each node is equipped with both classical and quantum computing resources. On the one hand, classical resources are composed of classical servers. Therefore, they can be efficiently deployed in a cluster format~\cite{anggara2024case, rak2021own, tong2018efficient}. On the other hand, quantum resources consist of Quantum Processing Units (QPUs)~\cite{ronkko2024premises} and their dedicated quantum control systems. From a system-wide perspective, the specific hardware implementations and parameter designs within the nodes remain completely transparent. Additionally, all nodes are interconnected via a network. Meanwhile, all nodes in this architecture are peer-to-peer and share an identical set of global control logic.

However, this novel architecture also exposes a systemic challenge. Due to the architectural complexity, existing programming interfaces struggle to handle it effectively. The specific challenge lies in how to express communication and synchronization between distributed classical processes and device-bound quantum execution.

\subsection{Related Work}
This requirement for hybrid computing differs fundamentally from traditional distributed quantum computing. The following analysis of related work aims to clarify the fundamental deficiencies of existing technologies when programming for large-scale classical-quantum heterogeneous hybrid task flows.

Distributed Quantum Computing (DQC)~\cite{caleffi2024distributed, barral2024review, diadamod2021distributed} is widely considered an effective approach to break through the scaling bottlenecks of single quantum processors and realize large-scale quantum computing. Its core idea lies in interconnecting multiple quantum nodes via quantum and classical communication channels. Recent review literature has systematically summarized the architectural models, network mechanisms, and software challenges of DQC. They indicate that the realization of distributed execution requires coordinated support across hardware, communication protocols, compilers, and programming environments~\cite{caleffi2024distributed}. Early foundational works (e.g., proxy graph state quantum computing) have further demonstrated the feasibility of utilizing networked quantum nodes to cooperatively achieve larger-scale computations, thereby establishing an important conceptual foundation for distributed quantum systems~\cite{benjamin2006brokered}. However, the aforementioned research primarily discusses DQC interconnection schemes at the architectural and physical network levels. They do not address the issue of a unified message communication interface oriented toward large-scale classical-quantum heterogeneous hybrid task flows. Therefore, directly applying existing DQC architectures to the heterogeneous hybrid programming scenarios targeted in this paper fails to resolve the technical challenges of jointly scheduling and coordinating classical and quantum resources under a unified communication model.

In the software abstraction ecosystem of existing DQC, design methodologies inspired by MPI occupy a significant position. For a long time, MPI has been the de facto standard programming model for classical distributed memory systems~\cite{mpi40_standard}. Based on this consensus, multiple studies have explored migrating communication-oriented abstraction concepts to the quantum domain: QMPI, proposed by Häner et al., became the first framework to utilize message-passing concepts to express distributed quantum computing~\cite{haner2021distributed_qmpi}; subsequently, Shi et al. proposed a reference implementation of a quantum message passing interface, providing practical support for the development and evaluation of QMPI-style programming models~\cite{shi2023reference_qmpi}; furthermore, NetQMPI, proposed by Cárdama and Peña, is a distributed quantum application programming framework for quantum networks built upon the NetQASM SDK~\cite{cardama2025netqmpi}. These studies collectively demonstrate that MPI-like abstraction designs provide a natural entry point for distributed quantum programming, showing significant advantages particularly in modeling communication, synchronization, and coordination among multiple quantum nodes. However, existing QMPI-like research solely focuses on distributed quantum execution itself, with its attention confined to quantum communication and remote quantum operations across networked quantum processors. They severely lack support for the tightly coupled requirements in classical-quantum heterogeneous hybrid systems—such as classical control, distributed data exchange, iterative orchestration, and dynamic invocation of quantum kernels~\cite{cardama2025netqmpi,haner2021distributed_qmpi,shi2023reference_qmpi}. Consequently, directly applying existing QMPI-like frameworks to our scenario still fails to resolve the core challenge of cooperatively scheduling classical and quantum resources under a unified message communication model.

Meanwhile, recent developments in quantum software platforms and programming toolchains have further enriched the technological ecosystem of distributed hybrid computing. For instance, frameworks like Qiskit provide full-stack support for the development, compilation, and execution of quantum programs, and support execution across heterogeneous backends~\cite{javadiabhari2024qiskit}; CUDA Quantum and its C++-based heterogeneous programming model focus on realizing integrated quantum-classical workflows~\cite{kim2023cudaq, mccaskey2021extending_cpp}; at the intermediate representation (IR) and language levels, OpenQASM 3 provides richer specifications covering classical control and timing awareness~\cite{cross2022openqasm3}, while QIR aims to achieve interoperable integration across diverse quantum software ecosystems~\cite{qir2024spec}. These works profoundly indicate that future quantum applications will inevitably be embedded into larger heterogeneous software stacks rather than executed in isolation. This aligns highly with the evolutionary trend of distributed hybrid computing. However, the core objective of such works is not to provide an MPI-style distributed communication abstraction. Therefore, they cannot achieve the complex orchestration of hybrid computing across multiple classical-quantum nodes. If these platforms and toolchains are directly applied to the large-scale heterogeneous hybrid task flow scenarios addressed in this paper, they similarly fail to solve the difficulty of coordinating multi-node cooperative computing under a unified message communication framework.

In summary, the works most closely related to ours are NetQMPI~\cite{cardama2025netqmpi} QMPI~\cite{haner2021distributed_qmpi} and its derivative projects~\cite{shi2023reference_qmpi}. All three draw inspiration from classical MPI~\cite{mpi40_standard} in an attempt to construct communication structures within distributed quantum environments. However, although these existing studies have explored the application of MPI-like abstractions in the quantum domain, classical parallel computing interfaces have neither considered the unique physical characteristics of quantum computing nor addressed the complex requirements of heterogeneous interactions. Moreover, the distributed coordination of quantum measurement and control systems fails to form an effective linkage with classical computing power scheduling, directly resulting in completely incompatible interfaces between classical and quantum domains. In heterogeneous hybrid architectures, the severe lack of quantum hardware awareness and quantum process synchronization mechanisms, combined with the technical bottlenecks faced by control sharing, constitute insurmountable obstacles. It is precisely due to the existence of these challenges that existing communication architectures fundamentally cannot meet the deep practical requirements of classical-quantum heterogeneous hybrid applications.

In contrast, the work in this paper does not treat distributed quantum computing as an isolated ultimate goal. Instead, this paper focuses on the practical requirements of distributed hybrid computing within classical-quantum heterogeneous systems. From this novel perspective, the quantum processor is explicitly defined as a computational accelerator. It is seamlessly embedded into a larger-scale distributed classical infrastructure. This fundamental shift in research focus dictates that abstractions inspired by MPI must provide more than just quantum message passing capabilities; more critically, they must support the deep integration of classical communication, the efficient processing of hybrid control flows, and full compatibility with existing toolchains. This paradigm shift from "isolated quantum execution" to "end-to-end hybrid computing" is of paramount importance for driving heterogeneous hybrid distributed applications toward ultimate practical utility.

\subsection{Contributions}
To address the deficiencies of the aforementioned related work, this paper designs the MPI-Q programming framework. This framework aims to achieve efficient coordination of large-scale classical-quantum heterogeneous hybrid workflows. Regarding its underlying implementation, MPI-Q is developed based on C and Python. Furthermore, it is seamlessly compatible with mainstream quantum programming frameworks such as Qiskit. In the subsequent sections, Section 3 and Section 4 will detail the heterogeneous hybrid optimization mechanisms of MPI-Q and its specific communication operations.

The specific contributions of this paper are summarized as follows:

\begin{itemize}
    \item Proposing a heterogeneous hybrid communication domain model: In classical-quantum heterogeneous systems, resource attributes differ significantly, and quantum tasks exhibit strong hardware dependencies. To address these characteristics, this paper defines a heterogeneous hybrid communicator model. This model not only achieves unified management of classical and quantum processes but also realizes the unified abstraction and precise isolation of heterogeneous resources. Meanwhile, it perfectly adapts to the differentiated resource mapping mechanisms of heterogeneous processes onto physical hardware.
    \item Constructing a heterogeneous lightweight communication architecture: The conventional "classical host relay" paradigm inevitably results in protracted data interaction chains and inefficient feedback. To overcome this bottleneck, this paper constructs a lightweight communication architecture based on localized Quantum MonitorProcesses. This architecture simplifies the multi-tier communication chain into a single-stage communication model. Consequently, the end-to-end communication path is significantly shortened, and the latency induced by secondary compilation at the target node is effectively circumvented.
    \item Designing a heterogeneous hybrid synchronization mechanism: In the context of heterogeneous hybrid distributed computing, multi-node quantum operations impose rigorous demands for precise timing control. To address this challenge, we established a heterogeneous hybrid synchronization mechanism. This mechanism defines unified synchronization primitives, reusing the classical MPI synchronization mechanism to ensure classical process coordination. Furthermore, by integrating Socket communication with hardware clock triggering techniques, it achieves precise timing alignment across distributed quantum monitor processes.
    \item Providing standardized programming support: MPI-Q enhances existing programming languages. It does this by adding message-passing capabilities between classical and quantum processes. Because of this enhancement, programmers can efficiently implement and deploy large-scale classical-quantum heterogeneous distributed computing applications.
\end{itemize}

\section{An Overview of MPI-Q}

MPI-Q is a message communication interface library dedicated to large-scale classical-quantum heterogeneous hybrid distributed computing. Its core design philosophy is to treat QPU as a heterogeneous accelerator, seamlessly embedding it into existing distributed classical computing infrastructures. MPI-Q chooses to deeply extend the classical MPI framework. This is because MPI, as the de facto standard in distributed computing, provides a natural entry point for the unified management of heterogeneous systems through its mature communicator abstraction and synchronization mechanisms. Through the deep synergy between "low-level optimization mechanisms" and "upper-layer communication interfaces," MPI-Q constructs a vertically integrated architecture spanning from underlying hardware resources directly to upper-layer programming applications. This architecture aims to comprehensively resolve the fundamental differences between classical and quantum systems across three dimensions: resource mapping, data interaction, and timing control. Ultimately, it provides developers with a standardized, efficient, and user-friendly programming support system.

\begin{figure*}[!ht]
    \centering
    \includegraphics[width=.8\linewidth]{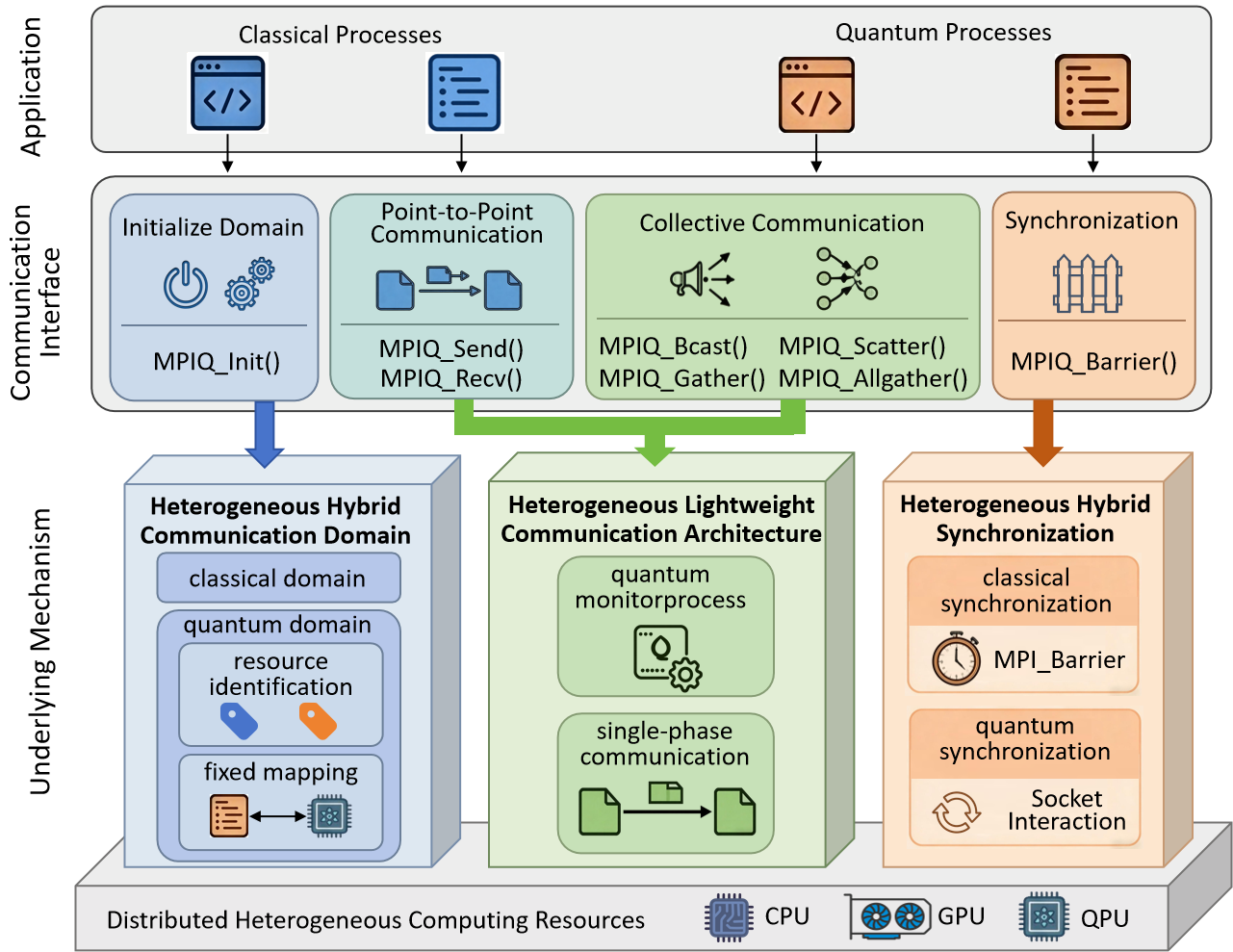}
    \caption{MPI-Q Overall Framework.}
    \label{fig:1}
\end{figure*}

The overall architectural design of MPI-Q is illustrated in Figure \ref{fig:1}. It consists of three core optimization mechanisms and a set of standardized communication interfaces. Crucially, these two major components are tightly coupled. As a result, they form a cohesive and organic whole.

The heterogeneous hybrid communicator model constitutes the cornerstone of resource management in the MPI-Q architecture. Specifically, this model extends the three-layer architecture of communicators in the classical computing domain to accommodate processes from the quantum computing domain. Consequently, it successfully achieves a unified abstraction of heterogeneous hybrid resources. Considering the fundamental differences in physical characteristics between classical and quantum computing, this model provides an innovative solution path. It addresses these differences by constructing a heterogeneous topology that includes both classical virtual processors and quantum virtual processors. This design perfectly adapts to different underlying resource mapping mechanisms. On the one hand, for classical computing resources, it supports random adaptive allocation through the virtual processor topology, thereby maintaining the flexibility of task scheduling. On the other hand, for quantum computing resources, due to their strong hardware dependency, the topology adapts to a strictly fixed mapping mechanism, thus establishing deterministic hardware binding pathways. In conclusion, this model effectively guarantees the precise scheduling and strict isolation management of heterogeneous resources.

The heterogeneous lightweight communication architecture serves as the robust data transmission pathway for MPI-Q. Thus, it guarantees the highly efficient execution of communication interfaces. This architecture introduces localized quantum MonitorProcesses. Traditionally, systems use a "classical host-machine relay" mode. However, this architecture simplifies it into a single-stage communication model. Specifically, the classical node directly connects to the quantum MonitorProcess. The control node completes quantum compilation and generates waveform data in advance. Subsequently, it transmits this data directly to the MonitorProcess on the target node. This design choice drastically shortens the data transmission path. As a result, it significantly reduces end-to-end communication latency. Furthermore, it enables upper-layer point-to-point and collective communication operations to achieve low-latency and high-bandwidth data interaction.

The heterogeneous hybrid synchronization mechanism is the core of timing control in the MPI-Q architecture. Crucially, it guarantees the strict timing consistency of communication operations. This mechanism is encapsulated through unified synchronization primitives. On one hand, it reuses classical MPI synchronization mechanisms. This ensures synergy among classical processes. On the other hand, it leverages Socket communication and hardware clock trigger technologies. Through this, it achieves precise timing alignment across multi-node quantum MonitorProcesses. In a distributed environment, this mechanism ensures that multi-node quantum operations meet stringent timing requirements. Ultimately, it provides critical support for the correctness of hybrid computations.

Supported by these three underlying mechanisms, MPI-Q explicitly addresses classical-quantum heterogeneous collaborative scenarios. To this end, it designs four major categories of standardized communication interface functions. These include communication context initialization, point-to-point communication, collective communication, and synchronization operations. Together, they form a comprehensive communication operation system. These interface functions are designed around the interaction needs between classical and quantum processes. Importantly, they maintain semantic consistency with classical MPI interfaces. In addition, they introduce exclusive capabilities. Specifically, these include quantum hardware addressing, quantum waveform data transmission, and multi-node quantum synergy. Consequently, they enable the rapid construction of hybrid communication domains and targeted data transmission. Furthermore, they facilitate batch data distribution, result aggregation, and hierarchical synchronization. Overall, they provide full-lifecycle communication support for task distribution, parallel execution, and result collection in hybrid computing. This set of interfaces is fully compatible with the classical MPI programming paradigm. Moreover, it effectively shields the complexity of the underlying heterogeneous hardware. Because of this, programmers can utilize existing programming languages (such as C and Python) to efficiently build large-scale hybrid applications.

Overall, the MPI-Q framework constructs a logically rigorous tri-fold architecture: it achieves the unified abstraction and management of heterogeneous processes through the heterogeneous hybrid communicator; it guarantees efficient cross-node data interaction within heterogeneous hybrid systems through the lightweight communication architecture; and it satisfies the strict requirements for multi-type timing synchronization in heterogeneous distributed computing scenarios through the heterogeneous hybrid synchronization mechanism. Building upon this foundation, MPI-Q relies on standardized interfaces to provide unified scheduling services to upper-layer applications. Ultimately, it successfully formulates a highly comprehensive, full-stack overall solution dedicated to classical-quantum hybrid computing.

\section{Heterogeneous Hybrid Optimization Mechanism}
Classical-quantum heterogeneous hybrid architectures impose strict system-level requirements. Specifically, they must achieve precise mapping between quantum processes and underlying quantum hardware. Furthermore, they require the ultra-low-latency transmission of both control data and measurement results. Finally, they must strictly guarantee the timing synchronization among distributed quantum processes. To address these complex challenges, this section details the core low-level designs of MPI-Q. Specifically, it proposes the heterogeneous hybrid communication domain model, the heterogeneous lightweight communication framework, and the heterogeneous synchronization mechanism. Through these three core mechanisms, MPI-Q specifically and effectively resolves all the aforementioned system-level synergy bottlenecks.

\subsection{Heterogeneous Hybrid Communication Domain}
In the field of distributed computing, a communication domain is a logical set composed of a group of processes. These processes can communicate with each other. Furthermore, this set provides an isolated communication space for these processes. In classical-quantum distributed hybrid computing, resource isolation and precise scheduling within heterogeneous systems are critically important. The reason is that this architecture involves vastly different task types, such as classical processes and quantum processes. Concurrently, it involves highly diverse hardware resources, including CPUs, GPUs, and quantum measurement and control devices. In addition, quantum tasks exhibit strong hardware dependencies. Therefore, they must be strictly bound to specific control devices and qubits for execution. If a unified communication management system is lacking, this heterogeneity can easily lead to incorrect mapping between tasks and resources.

To overcome this limitation, the MPI-Q framework innovatively proposes the concept of a heterogeneous hybrid communication domain. Based on this concept, researchers can strategically partition multiple hybrid communication domains within a hybrid system. These domains achieve mutual isolation at both physical and logical levels. However, the system can still achieve cross-domain control information sharing through a central controller. Specifically, the entire heterogeneous hybrid communication domain is strictly defined at the logical level by three core components. They are: the process group, the communication context, and the virtual processor topology~\cite{mpi40_standard, mpi40_standard, nikolopoulos_efficient_2018, hu_demystifying_2025}. These three components provide synergistic support to each other. Consequently, they collectively realize the logical abstraction and unified management of heterogeneous tasks and resources.

\begin{figure*}[!ht]
    \centering
    \includegraphics[width=.8\linewidth]{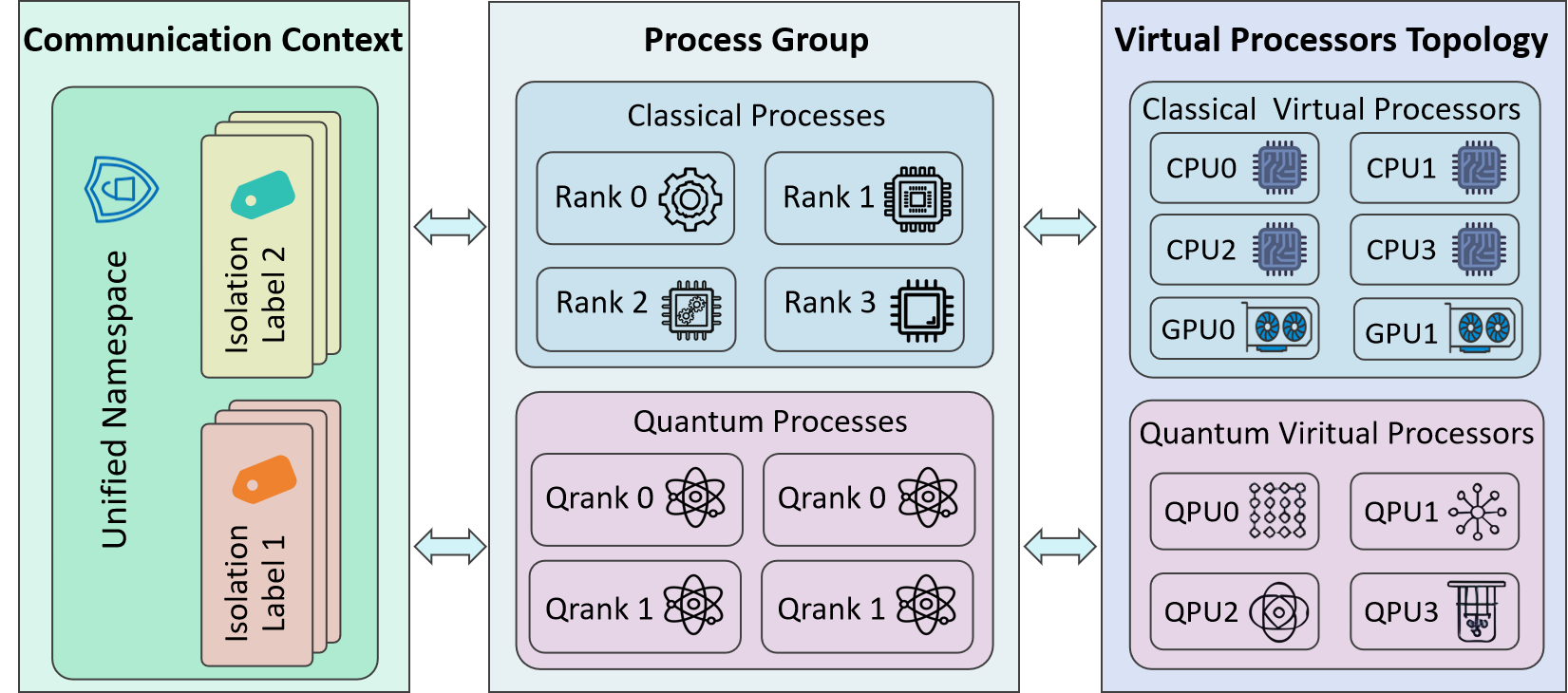}
    \caption{Hybrid Heterogeneous Communication Domain Model.}
    \label{fig:2}
\end{figure*}

Figure \ref{fig:2} clearly illustrates the overall model of the heterogeneous hybrid communication domain. First, the communication context provides strict security isolation tags and namespaces for message passing within the domain. As a result, it effectively prevents communication conflicts and data confusion among different hybrid communication domains when executing parallel tasks. Second, the process group acts as the main task executing body of the heterogeneous hybrid communication domain. This group is composed of both classical processes and quantum processes. Among them, classical processes use rank as their unique identifier, while quantum processes use qrank as their exclusive identifier. Subsequently, these two types of processes are partitioned into task units capable of parallel execution, based on the requirements of the hybrid computing tasks. Ultimately, they are jointly integrated into the global task scheduling system of the communication domain. Finally, the virtual processor topology is a logical abstraction of physical hardware resources. It consists of classical virtual processors and quantum virtual processors. Specifically, classical virtual processors correspond to the logical mapping of classical computing hardware. In contrast, quantum virtual processors build dedicated logical abstractions for quantum measurement and control devices as well as quantum processors. Through the design of this virtual processor topology, the system achieves a unified logical representation of heterogeneous physical hardware. Ultimately, this design successfully masks the underlying heterogeneity of the hardware.

Although the heterogeneous hybrid communication domain provides a unified logical abstraction of resources, the mapping from the virtual processor topology to the physical hardware must be resolved during actual execution. However, fundamental physical differences exist between classical and quantum computing. Therefore, this paper proposes distinct mapping mechanisms for these two types of virtual processors. Ultimately, this differentiated strategy establishes a deterministic pathway from logical abstraction to physical execution.

For mapping classical virtual processors to classical physical hardware (e.g., CPUs, GPUs), the system adopts a random adaptive allocation mechanism. This mechanism deliberately avoids establishing a fixed correspondence between them. Consequently, it maintains the flexibility and transparency of resource scheduling. Specifically, when a classical process initiates a resource request, the system first randomly selects a target node from the candidates. Subsequently, it verifies whether the node's performance and load status meet the current task requirements. If the conditions are met, the mapping is completed immediately. Otherwise, the system continues to iterate through other candidate nodes until an available resource is found. In short, by dynamically responding to task demands, this mechanism achieves highly efficient scheduling of classical hardware.

In contrast, for mapping quantum virtual processors to quantum physical hardware (e.g., control devices, QPUs), the system employs a strict fixed mapping allocation mechanism. This mechanism fundamentally aims to ensure the precise binding of quantum tasks to specific hardware. It consists of two core steps: resource identification and mapping binding. First, the system uses an $\{\text{IP}, \text{device\_id}\}$ tuple as the unique identifier for the hardware. Here, IP represents the physical network address of the quantum node, while device\_id represents the unique number of the control device within that node. Through static hardcoding, the system achieves precise hardware identification. Second, a dedicated mapping function uniquely associates the logical identifier of the quantum virtual processor with this tuple. Through this, the virtual processor is directionally bound to specific physical hardware. Concurrently, the system assigns an exclusive qrank to the corresponding quantum process. Ultimately, these operations establish a deterministic association: "quantum process - qrank - quantum virtual processor - quantum physical hardware". Therefore, this fixed mechanism ensures that each quantum process can only invoke its strictly designated hardware. This effectively prevents quantum state manipulation failures caused by resource mismatches.

In summary, the heterogeneous hybrid communication domain features a three-layer architecture comprising the process group, the communication context, and the virtual processor topology. Moreover, it integrates the aforementioned differentiated virtual-to-physical mapping strategies. Through these designs, the system successfully encapsulates complex hardware relationships and interaction rules within an abstraction layer. Consequently, developers only need to invoke unified function interfaces to accomplish task design and resource allocation at the logical level. This feature enables precise control and global coordination of heterogeneous resources without increasing programming complexity. Ultimately, this framework lays a solid foundation for resource management and communication scheduling in classical-quantum heterogeneous hybrid distributed computing.

\subsection{Heterogeneous Lightweight Communication Architecture}
\begin{figure*}[!ht]
    \centering
    \begin{subfigure}[b]{0.8\textwidth}
        \centering
        \includegraphics[width=\textwidth]{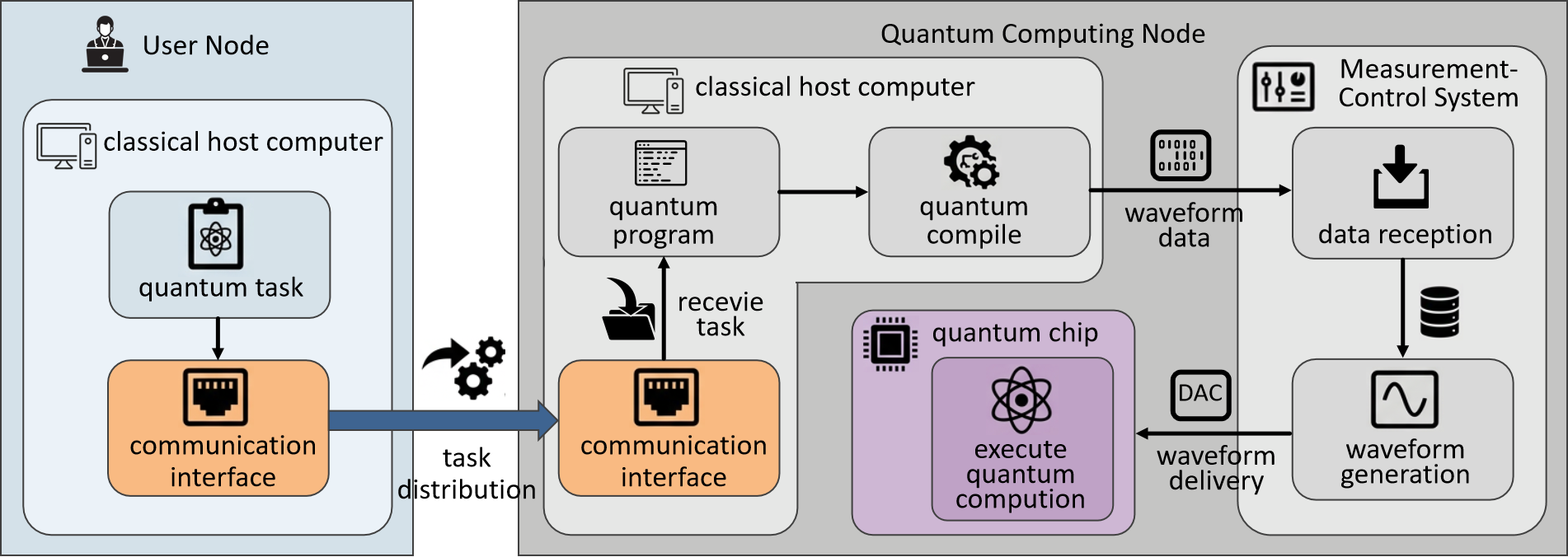}
        \caption{Traditional multi-stage communication link}
        \label{fig:3a}
    \end{subfigure}
    \hfill
    \begin{subfigure}[b]{0.8\textwidth}
        \centering
        \includegraphics[width=\textwidth]{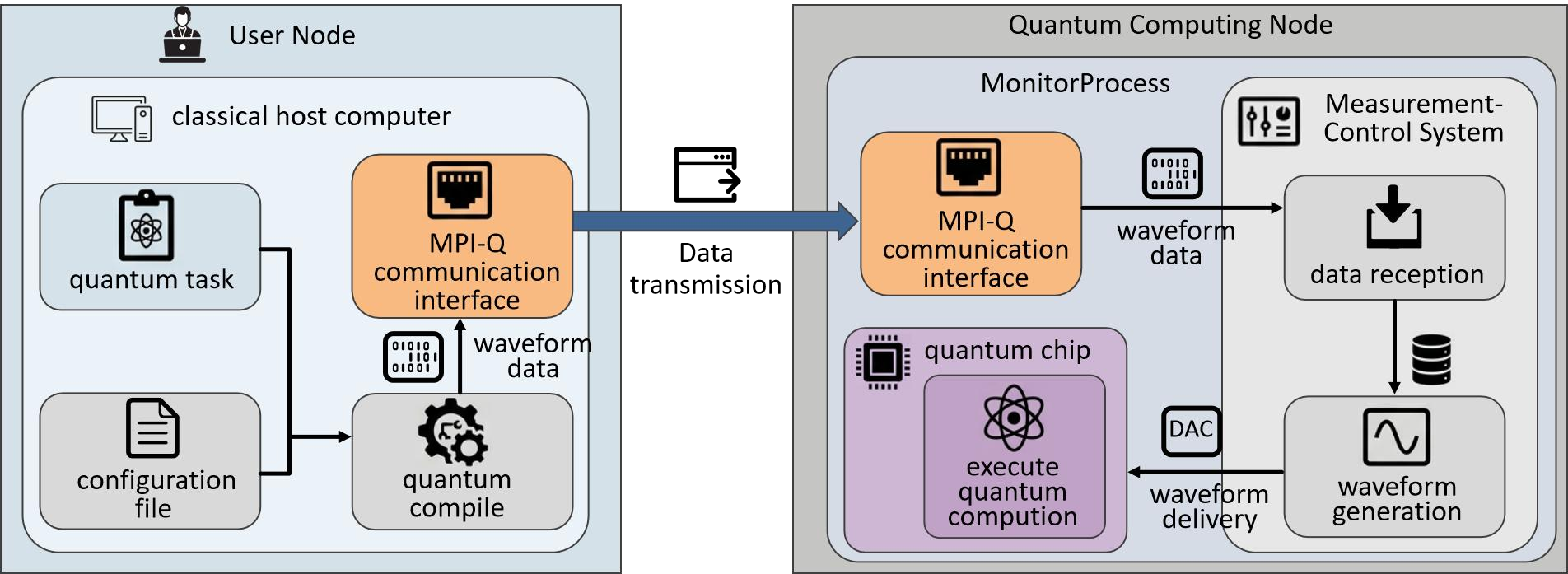}
        \caption{Lightweight communication link of MPI-Q}
        \label{fig:3b}
    \end{subfigure}
    \caption{Communication Architecture of Distributed Control in Heterogeneous Hybrid Systems}
    \label{fig:3}
\end{figure*}

When targeting large-scale classical-quantum heterogeneous distributed computing, employing a distributed control scheme is absolutely mandatory for multi-node synergy. However, traditional quantum computer systems typically adopt a highly centralized paradigm. Specifically, a classical host machine performs single-point scheduling of the quantum control system. Subsequently, the control system drives the quantum chip. Inherently, this mode only supports interaction between a single control terminal and a single quantum node. In a multi-quantum-node distributed framework based on this traditional mode, all control instructions and data severely depend on the centralized scheduling of the classical host. Consequently, massive data interactions in distributed hybrid computing must be relayed through this classical host. As illustrated in Figure \ref{fig:3a}, the traditional workflow is as follows:

\begin{enumerate}[label=\alph*)]
\item Task Distribution: The program on the user node splits the quantum computing tasks. Then, it distributes these tasks to the quantum computing nodes via traditional communication interfaces.
\item Local Compilation: After receiving the task, the quantum program on the computing node utilizes a local quantum compiler to generate hardware-dependent instructions. Next, it transmits these instructions to the quantum control system via the instruction dispatch unit.
\item Waveform Execution: The instruction receiving unit within the control system receives these instructions. Subsequently, it generates control data. Through a DAC (Digital-to-Analog Converter), this data is converted into analog waveforms. Finally, these waveform pulses are dispatched to the quantum chip via specific channels to execute the quantum computation.
\end{enumerate}

This traditional distributed control mode encompasses a lengthy, multi-stage process. It ranges from task processing and instruction generation on the classical host, to instruction conversion and dispatch by the control equipment, and finally to execution and result return by the quantum chip. In classical-quantum heterogeneous hybrid computing, the communication links of this control mode are excessively complex. As a direct result, the efficiency of control and feedback for cross-node heterogeneous resource collaborative scheduling is severely degraded.

To systematically resolve this, MPI-Q specifically designs a heterogeneous lightweight communication architecture tailored for hybrid distributed scenarios. Crucially, it introduces a MonitorProcess. This MonitorProcess uniformly manages the quantum control systems and quantum chips within each computing node. By doing so, it completely shatters the traditional paradigm of "receiving tasks first, then compiling locally." As illustrated in Figure \ref{fig:3b}, the optimized workflow is as follows:

\begin{enumerate}[label=\alph*)]
\item Pre-compilation and Transmission: The program on the user node splits the quantum tasks. Crucially, it completes the quantum compilation in advance, based on the system configuration files of the target quantum nodes. Subsequently, it sends the compiled waveform data directly to the MonitorProcess of the quantum computing node via the MPI-Q communication interface.
\item Direct Execution: Upon receiving the waveform data, the MonitorProcess on the quantum computing node directly invokes the quantum control system. It then immediately executes operations such as waveform generation and dispatch, ultimately driving the quantum computation.
\end{enumerate}

This architecture fundamentally relies on a lightweight communication framework for interaction. In stark contrast to the traditional multi-stage link (``classical host computer $\rightarrow$ classical host computer $\rightarrow$ quantum control system $\rightarrow$ quantum chip''), it dramatically simplifies the process into a single-stage communication model (``classical node $\rightarrow$ quantum monitor process''). At the user control end, the quantum circuit is pre-compiled based on the target's configuration. Therefore, it directly generates waveform data recognizable by the target system, entirely avoiding the secondary compilation latency at the target node. At the quantum computing end, the MonitorProcess achieves highly efficient management of the control system, swiftly handling waveform mapping and result measurement. When multiple nodes communicate based on this architecture, they can directly transmit qubit-level waveform data. This achieves precise transmission of the exact data required by the target control system. Ultimately, this architecture not only drastically shortens the end-to-end communication link but also significantly enhances the scalability of the entire system.

\subsection{Heterogeneous Hybrid Synchronization Mechanism}
\begin{figure}[!ht]
    \centering
    \includegraphics[width=.6\linewidth]{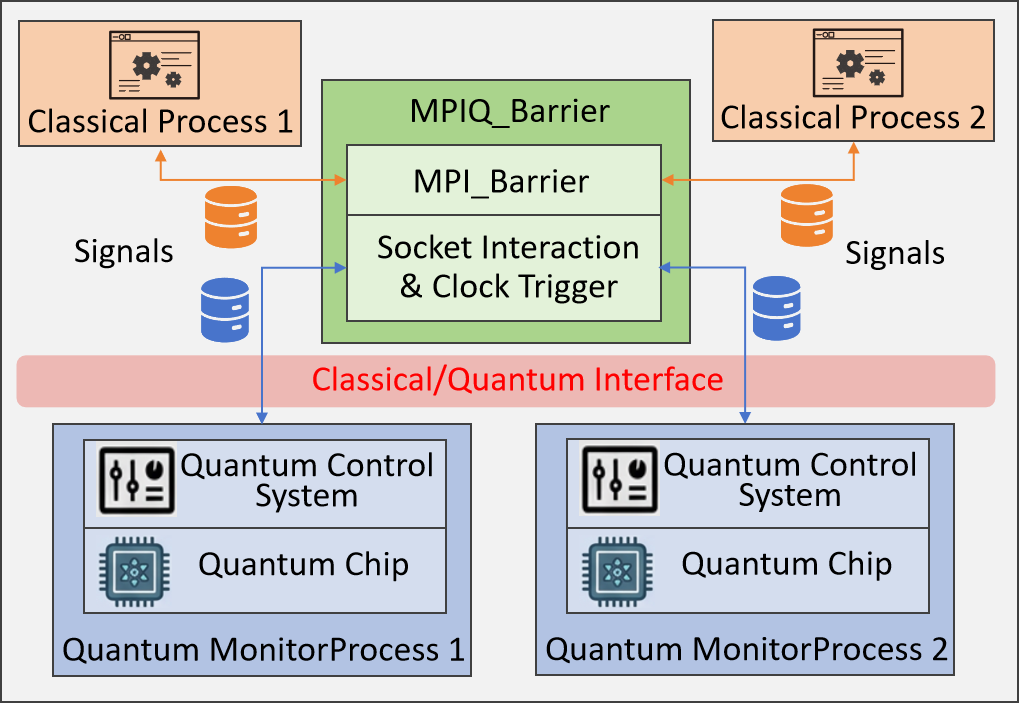}
    \caption{The Heterogeneous Hybrid Synchronization Mechanism.}
    \label{fig:4}
\end{figure}

Quantum control imposes exceptionally stringent requirements on timing precision. Specifically, any timing misalignment directly introduces severe control errors. In multi-node quantum networks, a high-precision synchronization mechanism must be employed to strictly unify the clocks and operational timing across all nodes. Only through this can the consistency of quantum state transmission, collaborative gate operations, and measurement results be guaranteed. However, traditional synchronization mechanisms inherently lack interfaces for interacting with quantum hardware, rendering efficient system coordination impossible. To address this critical gap, MPI-Q designs a heterogeneous hybrid synchronization mechanism, which is uniformly encapsulated and implemented by the MPIQ\_Barrier function. As illustrated in Figure \ref{fig:4}, this mechanism seamlessly achieves both synchronization among classical processes and synchronization among quantum MonitorProcesses.

Considering that MPI is a universally adopted distributed programming model in high-performance computing, its native synchronization mechanisms are highly mature for large-scale classical computation. Therefore, when handling classical process synchronization, MPIQ\_Barrier directly reuses the native MPI\_Barrier function.

Nevertheless, because the traditional MPI communication framework is incompatible with quantum processes, it cannot natively support synchronization among them. Consequently, the synchronization mechanism must be significantly extended for hybrid computing scenarios. For quantum processes, the core objective is to achieve the precise timing alignment of cross-node quantum operations. If different quantum nodes execute quantum circuits without completing strict synchronization, the timing deviations in multi-qubit gates will inevitably lead to a sharp decline in entanglement fidelity, ultimately invalidating the computational results. Because Socket is a mature cross-process communication technology capable of reliable, connection-oriented communication, MPI-Q designs a quantum process synchronization scheme based on the Socket mechanism. Through the MonitorProcess architecture, it achieves low-latency and highly reliable cross-node coordination. Specifically, each quantum MonitorProcess is responsible for managing both the local quantum computing resources and connecting to the classical interconnection network. The source process first obtains the $\{\text{IP}, \text{device\_id}\}$ physical identifier of the target process through the hybrid communication domain. Subsequently, it actively establishes a Socket connection and sends control flows, such as synchronization signals, to the target. Meanwhile, the target process listens on the designated TCP port of the classical node. Upon receiving the signal, it returns status information (e.g., synchronization readiness) to the source process.

\begin{algorithm}[htbp]
\caption{MPIQ\_Barrier Implementation}
\label{alg:mpiq_barrier}
\begin{algorithmic}[1]
\REQUIRE $flag$: Synchronization flag (0 for classical, 2 for quantum)
\ENSURE Achieve synchronization between classical or quantum processes

\STATE \textbf{\#define} CC $0$   \COMMENT{Between Classical Processes}
\STATE \textbf{\#define} QQ $2$  \COMMENT{Between Quantum MonitorProcess}

\IF{$flag == \text{CC}$}
    \STATE $\text{MPI\_Barrier}()$
    \COMMENT{Synchronization of Classical Processes}
\ELSIF{$flag == \text{QQ}$}
    \STATE $\text{Socket Interaction \& External Clock Synchronization}$
    \COMMENT{Synchronization of Quantum MonitorProcess}
\ENDIF
\end{algorithmic}
\end{algorithm}

Furthermore, from a hardware perspective, the heterogeneous hybrid system relies on high-stability reference clocks. The heterogeneous hybrid synchronization mechanism further realizes the precise timing alignment of quantum processes by deeply integrating with the quantum control system's hardware modules. These include clock calibration modules, delay measurement units, and dynamic compensation logic. By doing so, the triggering time deviation of quantum operations across different nodes can be strictly controlled within the allowable tolerance range of qubit coherence times and gate operation durations.

The specific implementation of the heterogeneous hybrid synchronization mechanism via the MPIQ\_Barrier function is detailed in Algorithm 1.

\section{Communication Operations}
MPI-Q's communication operations are deeply rooted in the robust framework of classical MPI communication primitives. Specifically, they are tailor-made communication interface functions explicitly designed for classical-quantum heterogeneous collaborative scenarios. To facilitate structured programming, these interfaces are systematically categorized into four core types: communication context initialization, point-to-point communication, collective communication, and synchronization operations.

\subsection{Initialization Operation}
The initialization functions are strictly utilized to achieve the initialization and configuration of the hybrid communication domain. Specifically, MPIQ\_Init serves as the primary initialization entry function for the MPI-Q framework. Fundamentally, it significantly extends the semantics of the classical MPI\_Init function. The execution flow of this function is highly structured. First, it invokes the classical MPI\_Init to initialize the classical communication domain. Subsequently, it reads the quantum node configuration files to initialize the quantum hardware resources. During this phase, it constructs the quantum communication domain and rigidly assigns a qrank identifier to each quantum process. Furthermore, it seamlessly integrates the classical and quantum communication domains to generate a unified hybrid communication domain. Simultaneously, it initializes critical functional modules, such as hardware-unified clock calibration and TCP Socket communication links. Finally, it launches a localized MonitorProcess on each quantum node, ultimately providing a unified heterogeneous distributed communication environment for the entire system..

\begin{figure}[H]
    \centering
    % First Row: 3 Figures
    \begin{minipage}[H]{0.3\textwidth}
        \centering
        \begin{subfigure}[H]{\textwidth}
            \includegraphics[width=\textwidth]{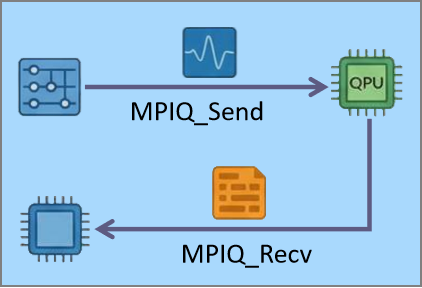}
            \caption{MPIQ\_Send/Rcev}
            \label{fig:5a}
        \end{subfigure}
    \end{minipage}
    \hfill
    \begin{minipage}[H]{0.3\textwidth}
        \centering
        \begin{subfigure}[H]{\textwidth}
            \includegraphics[width=\textwidth]{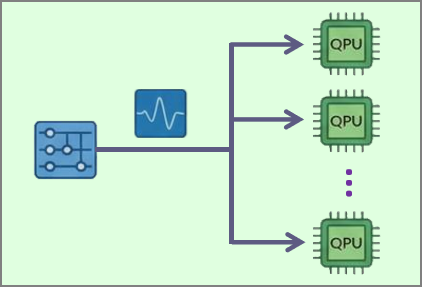}
            \caption{MPIQ\_Bcast}
            \label{fig:5b}
        \end{subfigure}
    \end{minipage}
    \hfill
    \begin{minipage}[H]{0.3\textwidth}
        \centering
        \begin{subfigure}[H]{\textwidth}
            \includegraphics[width=\textwidth]{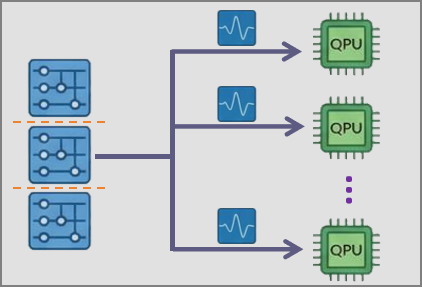}
            \caption{MPIQ\_Scatter}
            \label{fig:5c}
        \end{subfigure}
    \end{minipage}

    \vspace{1.5em}

    % Second Row: 3 Figures
    \begin{minipage}[H]{0.3\textwidth}
        \centering
        \begin{subfigure}[H]{\textwidth}
            \includegraphics[width=\textwidth]{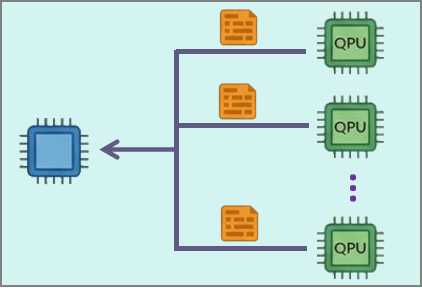}
            \caption{MPIQ\_Gather}
            \label{fig:5d}
        \end{subfigure}
    \end{minipage}
    \hfill
    \begin{minipage}[H]{0.3\textwidth}
        \centering
        \begin{subfigure}[H]{\textwidth}
            \includegraphics[width=\textwidth]{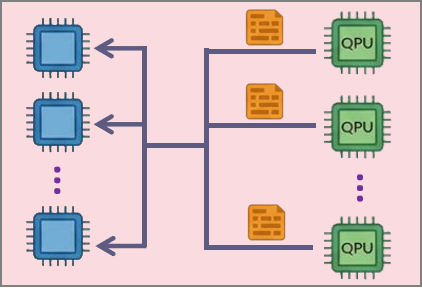}
            \caption{MPIQ\_Allgather}
            \label{fig:5e}
        \end{subfigure}
    \end{minipage}
    \hfill
    \begin{minipage}[H]{0.3\textwidth}
        \centering
        \begin{subfigure}[H]{\textwidth}
            \includegraphics[width=\textwidth]{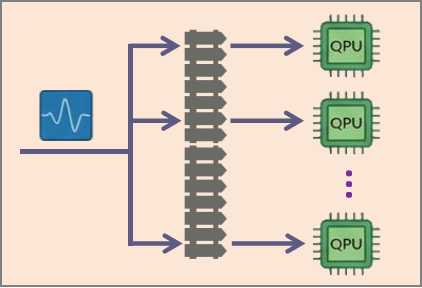}
            \caption{MPIQ\_Barrier}
            \label{fig:5f}
        \end{subfigure}
    \end{minipage}

    \vspace{1.0em}
    
    \begin{minipage}[H]{0.8\textwidth}
        \centering
        \begin{subfigure}[H]{\textwidth}
            \includegraphics[width=\textwidth]{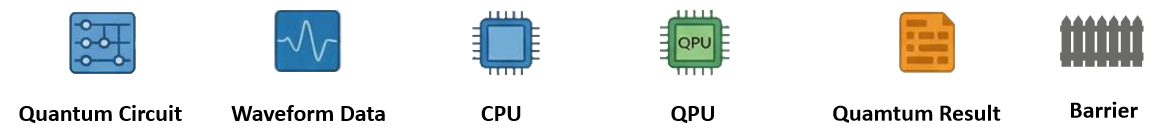}
            % \caption{}
        \end{subfigure}
    \end{minipage}

    \caption{MPI-Q Communication Operators.}
    \label{fig:5}
\end{figure}

\subsection{Point-to-Point Communication Operations}
Point-to-point communication constitutes the foundational communication operation of the MPI-Q communication architecture. Crucially, it heavily relies on a physical resource direct addressing mechanism to design the MPIQ\_Send and MPIQ\_Recv functions. As a result, it enables highly directional data transmission between classical computing nodes and specific quantum quantum control devices, as illustrated in Figure \ref{fig:5a}.

Drilling down into the details, the MPIQ\_Send function executes the directional dispatch of data from the classical process to the quantum MonitorProcess. To achieve precise targeting, it innovatively adopts the $\{\text{IP}, \text{device\_id}\}$ parameter pair to replace the traditional abstract rank. By leveraging the mapping relationships within the hybrid communication domain, it establishes a fixed connection from the classical process directly to the quantum control device. Moreover, it structurally stores the waveform data using a precise three-dimensional array. This array strictly follows a hierarchical "Compute Node - Quantum Control Device - Qubit" layout, perfectly adapting to the exact addressing requirements between data and hardware. Under the hood, intra-node communication employs shared memory zero-copy technology. This effectively circumvents the heavy overhead of kernel-mode context switching. Conversely, inter-node communication is extensively optimized based on the Socket protocol.

Correspondingly, the MPIQ\_Recv function handles the transmission of execution results back from the quantum MonitorProcess to the classical process. Strategically, its parameter design fully supports precise matching and dynamic adaptation for data reception. Consequently, it can effortlessly satisfy the result-return demands of varying-scale quantum tasks. As the direct complementary function to MPIQ\_Send, its underlying implementation principles remain strictly consistent with those of the send operation.

\subsection{Collective Communication Operations}
To further embody the collaborative optimization strategy for distributed quantum tasks, this paper develops a suite of collective communication operation interfaces for MPI-Q. These include MPIQ\_Bcast, MPIQ\_Scatter, MPIQ\_Gather, and MPIQ\_Allgather. Fundamentally, each collective operation can be flexibly implemented through various optimized point-to-point communication schemes.

Specifically, the MPIQ\_Bcast function achieves bulk data broadcasting from a classical process to multiple quantum processes, as illustrated in Figure \ref{fig:5b}. It is primarily utilized to broadcast qubit waveform data from the classical control node to all quantum control devices within the hybrid communication domain. Consequently, it is highly suitable for scenarios requiring multiple quantum nodes to synchronously execute identical quantum operations (e.g., distributed quantum entanglement state preparation). Its core advantage lies in enabling one-time data distribution coupled with synchronous multi-node reception.

Conversely, the MPIQ\_Scatter function facilitates the directional scattering of data from a classical process to multiple quantum processes, as shown in Figure \ref{fig:5c}. This function allows the classical control node to distribute qubit waveform data to specific quantum control devices according to pre-established mapping relationships. By utilizing a structured mapping array named send\_q, it achieves the precise decomposition of quantum circuit segments, as detailed in Algorithm 2.

\begin{algorithm}[htbp]
\caption{Send\_q Array Definition}
\label{alg:send_q}
\begin{algorithmic}[1]
\REQUIRE Qubit grouping information
\ENSURE Qubit-device mapping array $\text{send\_q}$

\STATE $\text{send\_q} = \{ $
\STATE \quad $\{\text{qubit\_0},\ \text{qubit\_1},\ \text{end}\},$
\STATE \quad $\{\text{qubit\_2},\ \text{qubit\_3},\ \text{end}\},$
\STATE \quad $\dots$
\STATE $\}$
\end{algorithmic}
\end{algorithm}

Here, send\_q defines the exact mapping between qubits and control systems, with each sub-array representing a specific quantum computational sub-task. During execution, the function parses this mapping and employs optimized point-to-point mechanisms for directional distribution. This makes it ideal for distributed split-execution scenarios (e.g., multi-node collaboration for complex circuits), ensuring precise matching between data and target hardware resources.

For data collection, the MPIQ\_Gather function realizes the aggregation of results from multiple quantum processes to a single classical process, as depicted in Figure \ref{fig:5d}. It enables the classical node to collect waveform data or measurement results from all quantum nodes in the network. Therefore, it is perfectly suited for result-summarization scenarios (e.g., centralized processing of multi-node measurements), offering unified aggregation and structural integration of multi-source data.

Finally, the MPIQ\_Allgather function expands upon this by achieving result aggregation from multiple quantum processes to multiple classical processes, as seen in Figure \ref{fig:5e}. Architecturally, it adopts a two-tier "Collect + Distribute" framework. First, the master node invokes MPIQ\_Gather to aggregate the full set of quantum data. Subsequently, it utilizes the classical MPI\_Allgather to distribute this aggregated data to all classical processes. This operation is indispensable for scenarios where multiple classical nodes collaboratively process large-scale quantum tasks (e.g., the synergy between distributed classical optimization algorithms and quantum computing), as it guarantees global sharing and collaborative processing of quantum data across the classical cluster.

\subsection{Synchronization Operations}
Synchronization is the critical function for achieving high-precision timing coordination among multiple quantum processes, as illustrated in Figure \ref{fig:5f}. Specifically, the MPIQ\_Barrier function serves as the collective synchronization primitive tailored for classical-quantum heterogeneous scenarios within the MPI-Q framework. Fundamentally, it inherits the strict synchronization semantics of classical MPI. It not only supports synchronization among classical processes but also enables coordinated synchronization among quantum processes via TCP Socket communication mechanisms. Consequently, this function ensures synchronization consistency on the classical side while effectively resolving timing alignment issues on the quantum side. It acts as the linchpin technology supporting the timing consistency of classical-quantum hybrid algorithms. The detailed implementation of the MPIQ\_Barrier function has been thoroughly elaborated in Section 3.3.

In summary, through the deep collaboration of the aforementioned four core modules, the MPI-Q communication architecture successfully constructs standardized interfaces perfectly adapted for classical-quantum heterogeneous distributed computing. It perfectly embodies the design philosophies of precise heterogeneous resource adaptation, low-latency communication, and high-precision timing synchronization. Ultimately, it provides a unified, highly efficient, and exceptionally reliable programming framework for the realization of large-scale, distributed classical-quantum hybrid applications.

\section{Case Study}
This section aims to validate the practical effectiveness of MPI-Q in large-scale classical-quantum distributed computing. To achieve this, we design a two-phase case analysis encompassing functional verification and performance evaluation. First, in the functional verification phase, we use a typical distributed quantum computing task as a case study. This verifies the framework's comprehensive support for heterogeneous hybrid task flows. Second, in the performance evaluation phase, we focus on concurrency. We deploy large-scale node clusters for testing to deeply analyze MPI-Q's parallel execution efficiency and system scalability within heterogeneous environments.

Physically, the GHZ state is a fundamental carrier for multi-qubit global entanglement~\cite{galindo2002information}. Its preparation circuit depth scales linearly with the number of qubits~\cite{aktar2024scalable}. Currently, single-node hardware faces physical bottlenecks, such as limited qubit counts and extremely short decoherence times~\cite{hao2025optical, smith2022scaling}. Due to these limitations, a single quantum node can hardly prepare GHZ states of 100 qubits or more independently. However, circuit cutting technology provides a structural solution. It divides a massive GHZ preparation circuit into several sub-circuits and distributes them to different quantum nodes for parallel execution. Subsequently, the system utilizes classical communication networks to correlate the measurement results of each sub-circuit and reconstruct the global quantum state. This computing paradigm perfectly aligns with the underlying architectural characteristics of classical-quantum hybrid computing.

Based on this, our core evaluation task focuses on the distributed computing of large-scale GHZ circuits. We design two sets of controlled experiments: cutting granularity adaptability tests and node scalability tests. First, we compare the total execution time of MPI-Q against a traditional serial approach for the same task. This comparison quantitatively verifies the significant parallel speedup enabled by MPI-Q. Concurrently, we systematically test MPI-Q's performance across varying quantum node cluster sizes to verify its robust adaptability to large-scale heterogeneous clusters. Through these comprehensive evaluations, we aim to fully demonstrate MPI-Q's engineering practicality and outstanding performance in real-world distributed quantum computing scenarios.

\subsection{GHZ Quantum Circuit and Cutting Scheme}
\begin{figure}[H]
\centering
\includegraphics[width=0.5\textwidth]{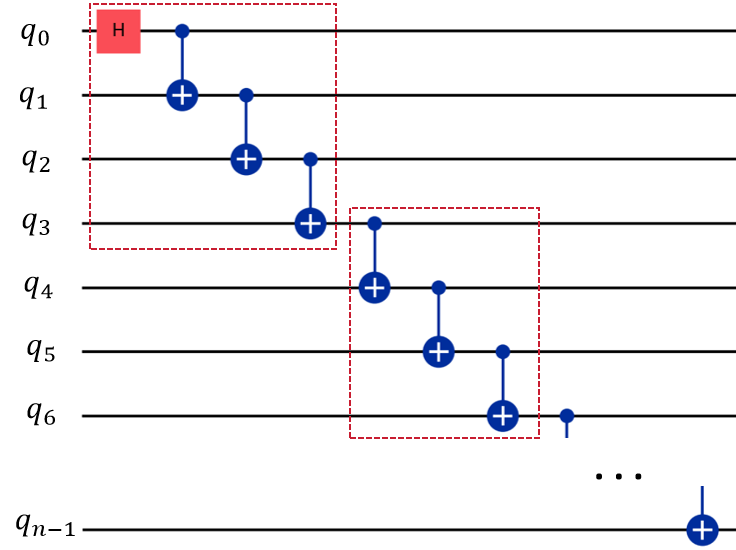}
\caption{GHZ Quantum Circuit and Circuit Cutting Scheme}
\label{fig:fig6}
\end{figure}

\textbf{GHZ Quantum Preparation Circuit.}
Figure \ref{fig:fig6} illustrates the quantum circuit for preparing an n qubit GHZ state, which consists of one Hadamard gate (H gate) and n-1 Controlled-NOT gates (CNOT gates)~\cite{galindo2002information, aktar2024scalable}. The core logic proceeds as follows: first, an H gate is applied to the initial qubit, placing it in a superposition of $|0\rangle$ and $|1\rangle$. Subsequently, CNOT gate operations are sequentially applied to the $i$-th and $i+1$ qubits ($i=0,2,...,n-2$) to establish multi-qubit entanglement. This ultimately generates the normalized GHZ entangled state: 
$|\text{GHZ}\rangle = \frac{1}{\sqrt{2}} (|00\dots0\rangle + |11\dots1\rangle)$.
With its concise structure and explicit entanglement properties, this circuit serves as an ideal test case for validating the parallel processing capabilities of distributed quantum computing frameworks.

\textbf{Equal-Granularity Cutting Scheme Based on Entanglement Edge Splitting.}
To enable the distributed parallel execution of large-scale GHZ circuits, this experiment employs an equal-granularity cutting algorithm based on entanglement edge splitting~\cite{bruss1999entanglement}, as shown in Figure \ref{fig:fig6}. The global GHZ preparation circuit, containing n qubits, is decomposed into m mutually independent sub-circuits. Each sub-circuit containS either $\lfloor n/m \rfloor$ or $\lceil n/m \rceil$ qubits and is uniquely mapped to a specific quantum node for execution. Crucially, this cutting scheme circumvents the need for direct, cross-node quantum entanglement interactions. Instead, it relies entirely on classical communication to correlate the execution results of the sub-circuits and reconstruct the global quantum state, thereby significantly reducing the physical implementation difficulty of distributed execution.

\subsection{Distributed Experimental Workflow}
Based on the communication interfaces of the MPI-Q framework, we design a distributed computing workflow tailored for the scalable preparation of large-scale GHZ states. This workflow constructs a collaborative execution pipeline between a classical control node and multiple quantum compute nodes, enabling the efficient decomposition, parallel execution, and result aggregation of massive quantum circuits. The entire pipeline, as illustrated in Figure \ref{fig:fig7}, is structurally divided into three phases. Each phase heavily relies on MPI-Q's communication primitives to achieve precise instruction dispatch, data transmission, and timing synchronization. The specific steps are detailed as follows:

\begin{figure}[H]
\centering
\includegraphics[width=1.0\textwidth]{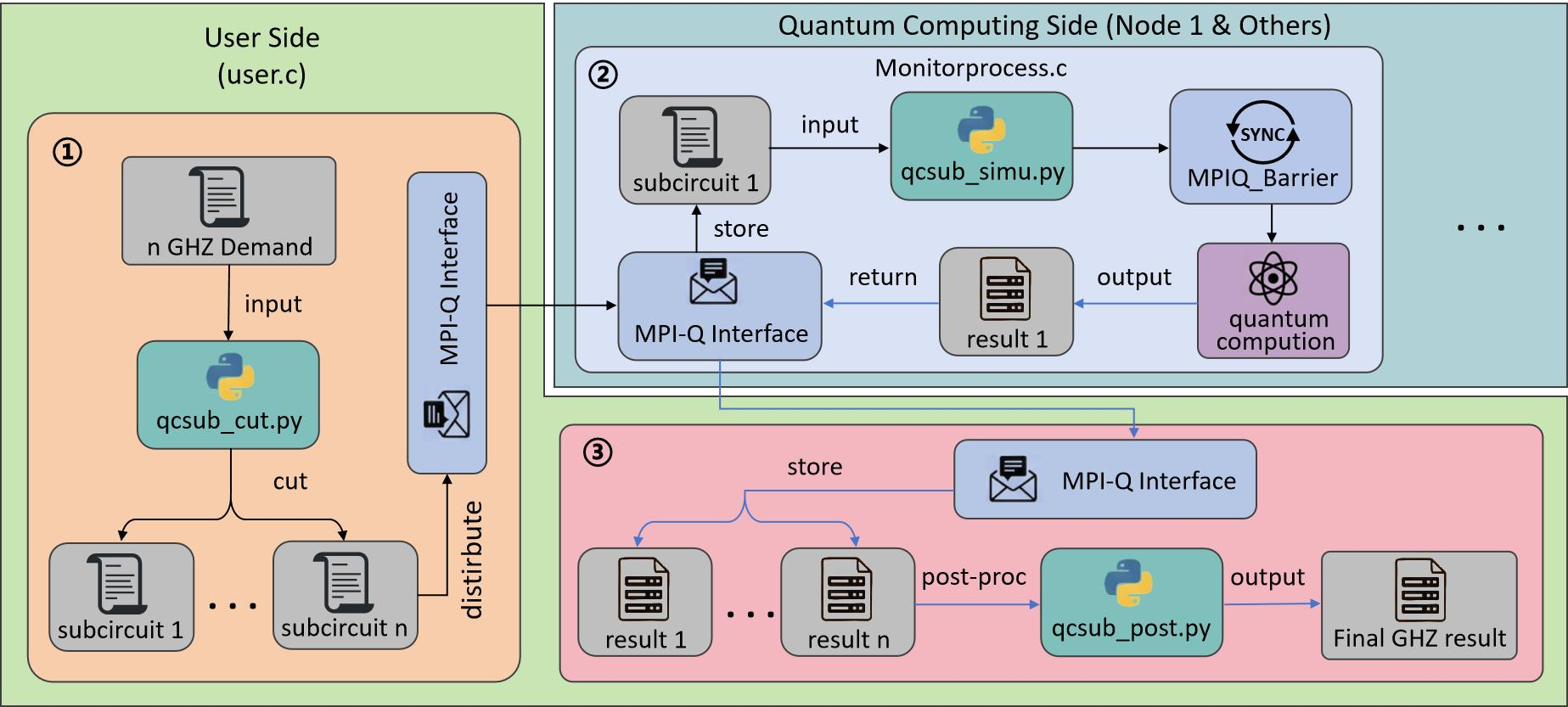}
\caption{Distributed Experiment Workflow}
\label{fig:fig7}
\end{figure}

\textbf{Task Initialization and Circuit Compilation}
The user submits a GHZ state preparation task via the \texttt{user.c} process on the classical control node, invoking the \texttt{qcsub\_cut.py} module for circuit cutting and compilation. Upon receiving the n-qubit task, this module employs a pre-defined equal-granularity cutting strategy to decompose the global circuit into m independent sub-circuits. In a quantum simulation environment, the compiler directly translates these sub-circuits into physical circuits. Conversely, for physical quantum systems, the sub-circuits must be compiled into directly recognizable waveform data, accommodating the specific configurations of the target nodes' quantum control systems. Finally, the sub-tasks are dispatched to pre-configured quantum nodes via MPI-Q communication interfaces.

\textbf{Parallel Execution of Quantum Sub-circuits}
The localized MonitorProcess, \texttt{Monitorprocess.c}, residing on each quantum compute node, accurately receives the assigned waveform data or compiled physical circuits via MPI-Q's reception interfaces and stores them locally. Subsequently, the \texttt{qcsub\_simu.py} module is invoked to read the sub-circuit data. Crucially, MPI-Q's synchronization interface (\texttt{MPIQ\_Barrier}) is utilized to enforce strict execution timing alignment across all quantum nodes, ensuring the synchronization of parallel execution. In simulated environments, nodes execute and measure sub-circuits synchronously via classical network interactions; on physical QPUs, the MonitorProcess integrates hardware clock triggering and latency calibration to further enhance multi-node timing precision. Upon completion, the measurement results are transmitted back to the classical control node via the \texttt{MPIQ\_Send} interface, closing the task processing loop for a single node. All nodes process sub-tasks concurrently following this unified workflow, significantly reducing the overall quantum execution time.

\textbf{Result Aggregation and GHZ State Reconstruction}
The classical control node leverages MPI-Q interfaces to receive the measurement results of sub-circuits returned by all quantum nodes, aggregating these results centrally. It then invokes the \texttt{qcsub\_post.py} module to reconstruct the sub-results based on the intrinsic entanglement properties of the GHZ state. Ultimately, the complete result of the n-qubit GHZ state preparation is generated and fed back to the \texttt{user.c} process for subsequent analysis and validation.

\section{Experiments}
\subsection{Experimental Setup}
The simulation experiments were conducted on a high-performance computing node at the Zhengzhou Supercomputing Center. This node is equipped with advanced computing hardware, high-version compilation environments, and a quantum simulation framework. For the purpose of this study, the node was configured to simulate a classical-quantum heterogeneous distributed cluster: a single CPU core was designated as the classical control node (responsible for circuit cutting, compilation/distribution, and result aggregation), while the remaining CPU cores simulated independent quantum computing nodes (responsible for executing quantum sub-circuits and returning measurement results). The specific software and hardware configurations are detailed in Table 1.

\begin{table}[htbp]
    \centering
    \caption{Experimental Software and Hardware Environment Configuration}
    \label{tab:env}
    \begin{tabular}{ll}
        \hline
        \textbf{Configuration Item} & \textbf{Version/Parameter} \\
        \hline
        CPU Cores & 32 \\
        CPU Model & Hygon C86 7185 (32-core Processor) \\
        Operating System & CentOS Linux release 7.9.2009 \\
        Programming Language & gcc 12.2.0, python 3.10.0 \\
        Quantum Simulator & pyqpanda 3.8.5, qiskit  0.42.0 \\
        \hline
    \end{tabular}
\end{table}

\subsection{Experimental Results and Analysis}
To comprehensively validate the technical advantages and practical characteristics of MPI-Q, the experimental evaluation is divided into two parts: Adaptability to Cutting Granularity and Node Scalability.

To quantitatively assess MPI-Q's performance, we designed the following two controlled experiments. The Adaptability to Cutting Granularity test fixes the number of quantum nodes to explore MPI-Q's adaptability and acceleration effects under varying sub-circuit cutting granularities. Conversely, the Node Scalability test fixes the sub-circuit scale to analyze MPI-Q's parallel execution efficiency and horizontal scaling capability across different numbers of quantum nodes. For these experiments, we define three key metrics: \textbf{Serial Execution Time} ($T_{\text{serial}}$) is the total time required for a single node to execute all sub-circuits sequentially; \textbf{Parallel Execution Time} ($T_{\text{parallel}}$) is the total time consumed by multiple nodes executing all sub-circuits concurrently under the MPI-Q framework; and \textbf{Speedup} ($S = T_{\text{serial}}/T_{\text{parallel}}$) quantifies the parallel acceleration effect achieved by MPI-Q.

\subsubsection{MPI-Q Cutting Granularity Adaptability Test}
In the first experiment, we fixed the number of quantum compute nodes at 10 and gradually increased the total number of qubits in the GHZ circuit, adjusting the sub-circuit cutting granularity accordingly. We measured the serial execution time, MPI-Q parallel execution time, and the resulting speedup under different granularities. The experimental results are detailed in Table 2.

\begin{table}[htbp]
    \centering
    \caption{MPI-Q Cutting Granularity Adaptability Test Results}
    \label{tab:granularity}
    \begin{tabular}{cccccc}
        \hline
        \textbf{GHZ Total} & \textbf{Quantum} & \textbf{Sub-circuit} & \textbf{Serial} & \textbf{MPI-Q Parallel} & \textbf{Speedup} \\
        \textbf{Circuit Size} & \textbf{Nodes} & \textbf{Size} & \textbf{Time (s)} & \textbf{Time (s)} & \\
        \hline
        40 & 10 & 4 & 13.29 & 2.57 & 5.18 \\
        80 & 10 & 8 & 14.11 & 2.67 & 5.29 \\
        120 & 10 & 12 & 14.73 & 2.76 & 5.33 \\
        160 & 10 & 16 & 18.72 & 3.14 & 5.97 \\
        200 & 10 & 20 & 63.52 & 8.82 & 7.20 \\
        210 & 10 & 21 & 119.00 & 15.22 & 7.82 \\
        220 & 10 & 22 & 227.49 & 28.85 & 7.89 \\
        230 & 10 & 23 & 465.35 & 55.96 & 8.32 \\
        240 & 10 & 24 & 948.93 & 111.25 & 8.53 \\
        250 & 10 & 25 & 1958.48 & 234.16 & 8.36 \\
        \hline
    \end{tabular}
\end{table}

\begin{figure}[H]
\centering
\includegraphics[width=0.8\textwidth]{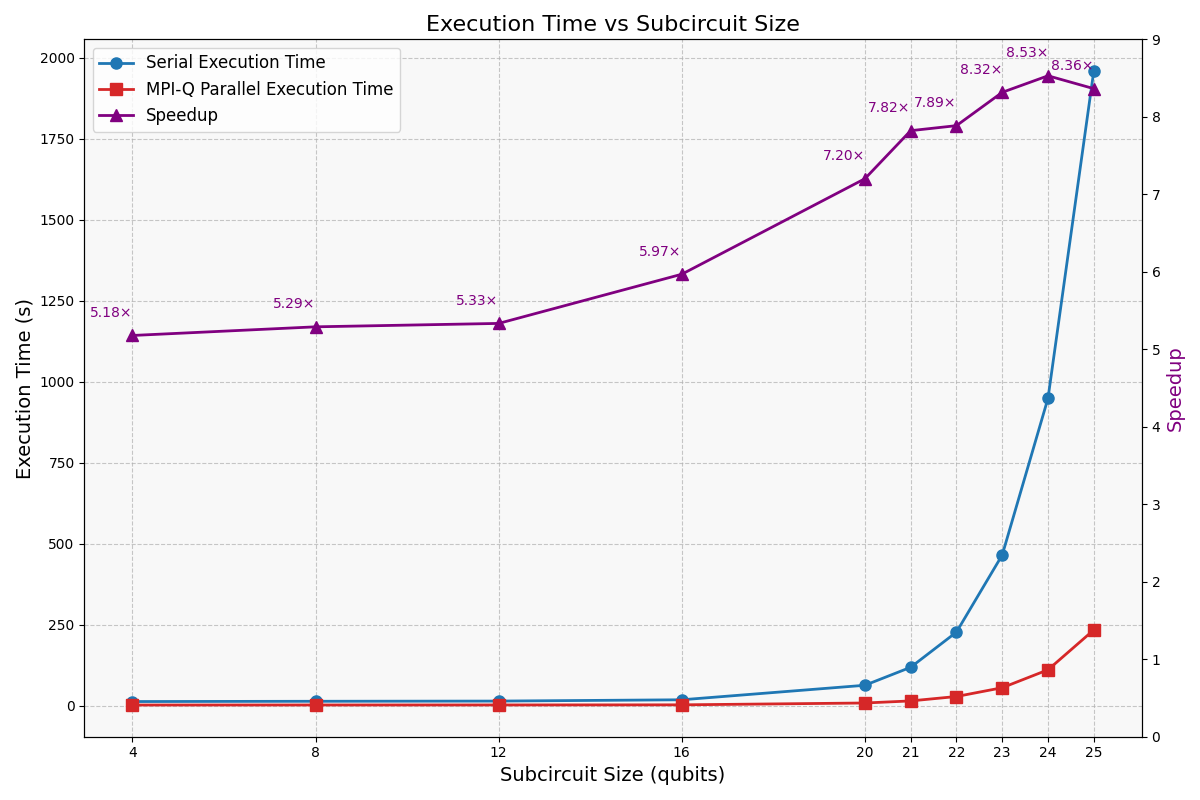}
\caption{MPI-Q Cutting Granularity Adaptability Test: Serial/Parallel Execution Time and Speedup vs. GHZ Circuit Scale}
\label{fig:fig8}
\end{figure}

As illustrated in Figure \ref{fig:fig8}, the speedup of MPI-Q parallel computation exhibits a trend of initial growth followed by a plateau as the cutting granularity increases:
\begin{itemize}
    \item When the total GHZ circuit scale is under 200 qubits: The sub-circuit granularity is relatively small. The dominant time cost of the task is concentrated on data transmission between classical and quantum processes (communication-bound). Consequently, the parallel speedup increases slowly with granularity.
    \item When the total GHZ circuit scale reaches 200~230 qubits: The quantum computation time of the sub-circuits overtakes communication overhead to become the primary time bottleneck of the global task (computation-bound). In this phase, the parallel speedup increases significantly with granularity.
    \item When the total GHZ circuit scale exceeds 230 qubits: As the sub-circuit granularity further increases, the local computational load on each quantum node intensifies. Restricted by the finite number of available quantum nodes, the parallel speedup gradually converges to a stable plateau.
\end{itemize}

These results indicate that MPI-Q possesses excellent adaptability to various cutting granularities. It empowers users to flexibly tune the GHZ circuit cutting granularity based on actual quantum hardware resources and task requirements, thereby achieving an optimal balance between execution efficiency and resource utilization.

\subsubsection{MPI-Q Node Scalability Performance Test}
In the second experiment, we evaluated the node scalability of MPI-Q by fixing the sub-circuit scale at 20 qubits and gradually increasing the number of quantum compute nodes (ranging from 1 to 24). Correspondingly, the total scale of the GHZ circuit was scaled up. We measured the serial execution time, MPI-Q parallel execution time, and speedup under various node configurations, with the results summarized in Table 3.

\begin{table}[htbp]
    \centering
    \caption{MPI-Q Node Scalability Performance Test Results}
    \label{tab:scalability}
    \begin{tabular}{cccccc}
        \hline
        \textbf{GHZ Total} & \textbf{Quantum} & \textbf{Sub-circuit} & \textbf{Serial} & \textbf{MPI-Q Parallel} & \textbf{Speedup} \\
        \textbf{Circuit Size} & \textbf{Nodes} & \textbf{Size} & \textbf{Time (s)} & \textbf{Time (s)} & \\
        \hline
        20 & 1 & 20 & 0.89 & 0.92 & 0.97 \\
        40 & 2 & 20 & 10.41 & 9.59 & 1.09 \\
        80 & 4 & 20 & 17.99 & 8.78 & 2.05 \\
        120 & 6 & 20 & 31.30 & 8.79 & 3.56 \\
        160 & 8 & 20 & 48.86 & 8.84 & 5.52 \\
        200 & 10 & 20 & 64.74 & 8.82 & 7.34 \\
        240 & 12 & 20 & 79.02 & 8.63 & 9.16 \\
        280 & 14 & 20 & 96.25 & 9.17 & 10.49 \\
        320 & 16 & 20 & 112.60 & 9.08 & 12.40 \\
        360 & 18 & 20 & 128.35 & 8.99 & 14.27 \\
        400 & 20 & 20 & 142.57 & 9.33 & 15.27 \\
        440 & 22 & 20 & 158.96 & 9.27 & 17.15 \\
        480 & 24 & 20 & 177.74 & 9.47 & 18.76 \\
        \hline
    \end{tabular}
\end{table}

\begin{figure}[!ht]
\centering
\includegraphics[width=0.8\textwidth]{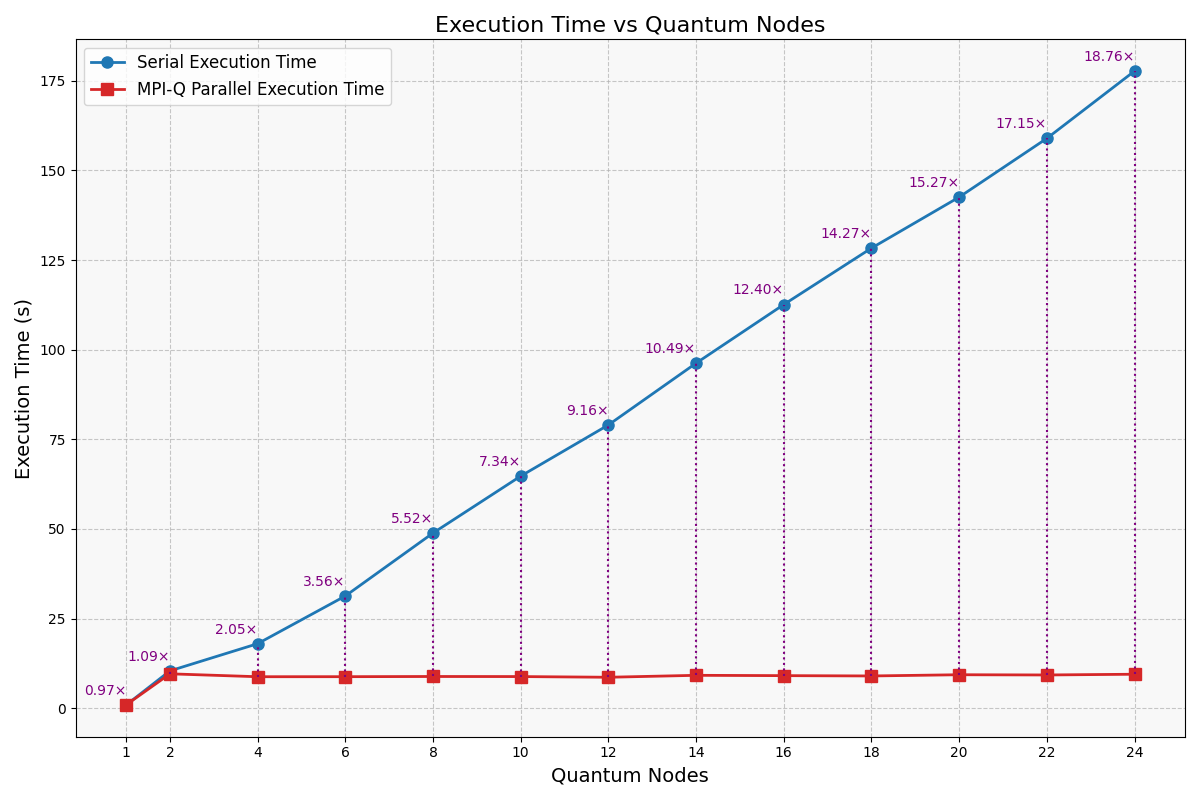}
\caption{MPI-Q Node Scalability Performance Test. Serial/Parallel Execution Time and Speedup vs. Quantum Node Count.}
\label{fig:fig9}
\end{figure}

As illustrated in Figure \ref{fig:fig9}, the parallel execution time and speedup of MPI-Q demonstrate a stable improvement trend as the number of quantum nodes increases. Notably, small-scale node configurations (1 and 2 nodes) exhibit unique performance characteristics that contrast sharply with large-scale configurations ($\geq$4 nodes). As the system scales from 4 to 24 nodes, the system exhibits excellent scalability:

\begin{itemize}
    \item Anomalies in Small-Scale Configurations: when the number of quantum nodes is 1, the system speedup is 0.97, and its parallel execution time (0.915 s) is slightly higher than the purely serial execution time (0.886 s). The core reason is that a single node lacks the physical hardware foundation for parallel computing. The task remains essentially fully serial, and the additional time overhead stems from process scheduling. For a 2-node configuration, the speedup is merely 1.09, showing negligible parallel acceleration. Due to the mechanics of circuit cutting, uneven sub-circuit sizes can occur, and the overall parallel time is bottlenecked by the execution of the largest sub-circuit.
    \item Stability of Parallel Execution Time: As the number of quantum nodes scales from 4 to 24—corresponding to an increase in the total GHZ circuit scale from 80 to 480 qubits—the total parallel execution time of MPI-Q remains exceptionally stable within a narrow margin of 8.6 to 9.5 seconds. The fluctuation is minimal and does not exhibit significant growth despite the massive expansion in task scale and node count.
    \item Near-Linear Growth in Speedup: The MPI-Q parallel speedup demonstrates a near-ideal linear growth trend with the addition of nodes. Scaling from 4 to 24 nodes, the speedup steadily climbs from 2.05$\times$ to 18.76$\times$, achieving a nearly 9-fold acceleration gain without showing obvious signs of performance saturation at 24 nodes.

\end{itemize}

These experimental results indicate that while MPI-Q does not provide a parallel advantage in minimal-scale scenarios (due to scheduling overhead and load imbalance), it possesses outstanding horizontal scalability (scale-out capability) in large-scale quantum node clusters. For configurations of 4 nodes or more, given a fixed single-node computational load, MPI-Q achieves ultra-stable parallel execution times and near-linear speedup. This proves that MPI-Q can seamlessly adapt to larger-scale clusters, effectively supporting the distributed and efficient execution of massive quantum circuits ranging from hundreds to potentially thousands of qubits.

\subsection{Experiment Summary}
Based on the experimental validation of distributed GHZ state preparation, MPI-Q's heterogeneous mixed communication domain model successfully realizes precise addressing and resource binding between classical and quantum processes. Its lightweight communication architecture significantly reduces the link latency associated with task dispatch and result retrieval. Furthermore, the heterogeneous mixed synchronization mechanism rigorously guarantees the strict timing consistency required for parallel sub-circuit execution across multiple quantum nodes. This experiment not only fully verifies MPI-Q's highly efficient adaptability for the distributed execution of massive quantum circuits, but also establishes a scalable engineering framework and a valuable experimental reference for diverse distributed quantum computing applications beyond GHZ state preparation.

\section{Conclusions}
In this paper, we propose and design MPI-Q, a unified message-passing interface library oriented towards large-scale classical-quantum distributed computing. Through three core architectural innovations—the heterogeneous mixed communication domain model, the heterogeneous lightweight communication architecture, and the heterogeneous mixed synchronization mechanism—MPI-Q effectively breaks through the technical bottlenecks of existing solutions regarding resource mapping, control sharing, and synchronization precision. By seamlessly introducing message communication capabilities between classical and quantum processes into existing programming languages, MPI-Q significantly enhances the collaborative efficiency of classical and quantum computing, thereby lowering the development threshold and system integration costs for hybrid applications. Our application validation results conclusively demonstrate that MPI-Q substantially improves the execution efficiency and scalability of classical-quantum distributed computing, providing crucial fundamental technical support for the construction of next-generation, large-scale hybrid computing systems.

\section*{Acknowledgment}
% This work was supported by National Key Research and Development Program of China (Grant No. 2023YFB4502500)
This work was supported by the National Key Research and Development Program of China under Grant 2023YFB4502500, and in part by the Public Service Platform for Quantum Computing Benchmarking.

% % 参考文献（arXiv兼容natbib，保留原样式）
% \bibliographystyle{cas-model2-names}
% \bibliography{cas-refs}
% \bibliographystyle{plain}
\bibliographystyle{unsrt}
\bibliography{references}

\begin{thebibliography}{10}

\bibitem{kumar2025secure}
Avinash Kumar.
\newblock Secure multi-programming for quantum computers.
\newblock Master's thesis, The University of Texas at Austin, Austin, TX, USA, May 2025.

\bibitem{cicero_simulation_2025}
Alessio Cicero, Mohammad~Ali Maleki, Muhammad~Waqar Azhar, and Anton Frisk.
\newblock Simulation of quantum computers: Review and acceleration opportunities.
\newblock {\em ACM Transactions on Quantum Computing}, 7(1), 2025.

\bibitem{singh_role_2023}
Suneet Singh.
\newblock The role of quantum superposition in computing.
\newblock {\em International Journal of Research Publication and Reviews}, 6(3):2889--2892, 2023.

\bibitem{shor1994algorithms}
Peter~W. Shor.
\newblock Algorithms for quantum computation: discrete logarithms and factoring.
\newblock In {\em Proceedings 35th Annual Symposium on Foundations of Computer Science}, pages 124--134. IEEE, 1994.

\bibitem{herrmann2024quantum}
Jan-R. Herrmann et~al.
\newblock The quantum-{Generative} {AI} symbiosis: Accelerating progress through technology.
\newblock {\em arXiv preprint arXiv:2408.xxxxx}, 2024.

\bibitem{sivak2022quantum}
William Sivak, Ian~D. Kivlichan, Yu~Cao, Craig Gidney, Margaret Martonosi, Jay~M. Gambetta, Dave endres, Ian~D. Kivlichan, Yu~Cao, Craig Gidney, Margaret Martonosi, Jay~M. Gambetta, and Dave endres.
\newblock Assessing requirements to scale to practical quantum advantage.
\newblock {\em arXiv preprint arXiv:2211.07629}, 11 2022.

\bibitem{mohseni2024how}
M.~Mohseni, A.~Scherer, K.~G. Johnson, O.~Wertheim, M.~Otten, N.~A. Aadit, Y.~Alexeev, K.~M. Bresniker, K.~Y. Camsari, B.~Chapman, S.~Chatterjee, G.~A. Dagnew, A.~Esposito, F.~Fahim, M.~Fiorentino, A.~Gajjar, A.~Khalid, X.~Kong, B.~Kulchytskyy, E.~Kyoseva, R.~Li, P.~A. Lott, I.~L. Markov, R.~F. McDermott, G.~Pedretti, P.~Rao, E.~Rieffel, A.~Silva, J.~Sorebo, P.~Spentzouris, Z.~Steiner, B.~Torosov, D.~Venturelli, R.~J. Visser, Z.~Webb, X.~Zhan, Y.~Cohen, P.~Ronagh, A.~Ho, R.~G. Beausoleil, and J.~M. Martinis.
\newblock How to build a quantum supercomputer: Scaling from hundreds to millions of qubits.
\newblock {\em arXiv preprint arXiv:2411.10406}, 2024.

\bibitem{gavande2024tackling}
S.~Gavande, B.~Nagappan, S.~Seo, and W.~K. Lee.
\newblock Tackling the challenges of adding pulse-level support to a quantum computing framework.
\newblock {\em PRX Quantum}, 6(1):010101, 2025.

\bibitem{saraiva2022materials}
A.~Saraiva, W.~H. Lim, C.~H. Yang, C.~C. Escott, and A.~Laucht.
\newblock Materials for silicon quantum dots and their impact on electron spin qubits.
\newblock {\em Advanced Functional Materials}, 32(3):2105488, 2022.

\bibitem{bernien2020probing}
H.~Bernien, B.~Hensen, W.~Pfaff, G.~Koolstra, M.~S. Blok, L.~Robledo, T.~H. Taminiau, M.~Markham, D.~J. Twitchen, V.~V. Dobrovitski, and R.~Hanson.
\newblock Probing many-body dynamics on a 51-atom quantum simulator.
\newblock {\em Physical Review Letters}, 124(11):110501, 2020.

\bibitem{hao2025optical}
X.~Hao, E.~S. Allgeyer, D.~R. Lee, D.~I. Schuster, S.~Choi, C.~Kim, A.~Somoroff, J.~Zhang, J.~Ye, and V.~Vuleti{\'c}.
\newblock Optical addressing enables a new architecture for spatially resolved quantum simulation.
\newblock {\em Nature}, 639(7989):596--601, 2025.

\bibitem{dong2021fast}
Yang Dong, Ce~Feng, Yu~Zheng, Xiang-Dong Chen, Guang-Can Guo, and Fang-Wen Sun.
\newblock Fast high-fidelity geometric quantum control with quantum brachistochrones.
\newblock {\em Physical Review Research}, 3(4):043177, 2021.

\bibitem{coelloperez2022quantum}
Eduardo~A. Coello~Pérez, Joey Bonitati, Dean Lee, Sofia Quaglioni, and Kyle~A. Wendt.
\newblock Quantum state preparation by adiabatic evolution with custom gates.
\newblock {\em Physical Review A}, 105(3):032403, 2022.

\bibitem{motzoi2009simple}
F.~Motzoi, J.~M. Gambetta, P.~Rebentrost, and F.~K. Wilhelm.
\newblock Simple pulses for elimination of leakage in weakly nonlinear qubits.
\newblock {\em Physical Review Letters}, 103(11):110501, 2009.

\bibitem{grumbling2019quantum}
Emily Grumbling and Mark Horowitz.
\newblock {\em Quantum Computing: Progress and Prospects}.
\newblock The National Academies Press, Washington, DC, 2019.

\bibitem{ding2020systematic}
Yongshan Ding, Pranav Gokhale, Sophia~Fuhui Lin, Richard Rines, Thomas Propson, and Margaret Martonosi.
\newblock Systematic crosstalk mitigation for superconducting qubits.
\newblock In {\em 2020 53rd Annual IEEE/ACM International Symposium on Microarchitecture (MICRO)}, pages 1001--1016, 2020.

\bibitem{smith2022scaling}
Kaitlin~N. Smith, Gokul Subramanian~Ravi, Jonathan~M. Baker, and Frederic~T. Chong.
\newblock Scaling superconducting quantum computers with chiplet architectures.
\newblock In {\em 2022 55th IEEE/ACM International Symposium on Microarchitecture (MICRO)}, pages 1092--1109, Chicago, IL, USA, oct 2022. IEEE, IEEE.
\newblock Also available as arXiv:2210.10921.

\bibitem{zappin2025quantum}
Jake Zappin et~al.
\newblock When quantum meets classical: Characterizing hybrid quantum-classical issues discussed in developer forums.
\newblock In {\em 2025 IEEE/ACM 47th International Conference on Software Engineering (ICSE)}, pages 2931--2943, 2025.

\bibitem{bensoussan2025taxonomy}
Avner Bensoussan, Gunel Jahangirova, and Mohammad~Reza Mousavi.
\newblock A taxonomy of real faults in hybrid quantum-classical architectures.
\newblock {\em ACM Transactions on Software Engineering and Methodology}, 2025.

\bibitem{mcclean2016theory}
Jarrod~R. McClean, Jonathan Romero, Ryan Babbush, and Alán Aspuru-Guzik.
\newblock The theory of variational hybrid quantum-classical algorithms.
\newblock {\em New Journal of Physics}, 18(2):023023, 2016.

\bibitem{esposito2025slurm}
Aniello Esposito and Utz-Uwe Haus.
\newblock Slurm heterogeneous jobs for hybrid classical-quantum workflows.
\newblock {\em arXiv preprint arXiv:2506.03846}, 2025.

\bibitem{chia2024hybrid}
Cleaven Chia, Ding Huang, Victor Leong, Jian~Feng Kong, and Kuan~Eng Johnson.
\newblock Hybrid quantum systems with artificial atoms in solid state.
\newblock {\em Advanced Quantum Technologies}, 7(5):2300461, 2024.

\bibitem{gazda2024pragma}
Arnaud Gazda and Océane Koska.
\newblock A pragma based c++ framework for hybrid quantum/classical computation.
\newblock {\em Science of Computer Programming}, 236:103119, 2024.

\bibitem{pasini2025enabling}
Matteo Pasini et~al.
\newblock Enabling hybrid quantum networks: Protocols and experimental progress toward memory-compatible links.
\newblock CEWQO 2025, 34th Central European Workshop on Quantum Optics, 2025.
\newblock Workshop contribution.

\bibitem{bravomontes2025architectural}
J.~A. Bravo-Montes, Miriam Bastante, and Cyril Allouche.
\newblock Architectural design and orchestration of heterogeneous quantum-classical computing systems.
\newblock In {\em Proceedings of the 1st International Conference on Quantum Computing and Engineering}, page 136536. SCITEPRESS, 2025.

\bibitem{mermin1990extreme}
N.~David Mermin.
\newblock Extreme quantum entanglement in a superposition of macroscopically distinct states.
\newblock {\em Physical Review Letters}, 65(15):1838--1840, 1990.

\bibitem{zurek2003decoherence}
Wojciech~Hubert Zurek.
\newblock Decoherence and the transition from quantum to classical -- revisited.
\newblock {\em Los Alamos Science}, 27:2--25, 2003.

\bibitem{miyadera2009nocloning}
Takayuki Miyadera and Hideki Imai.
\newblock No-cloning theorem on quantum logics.
\newblock {\em Journal of Mathematical Physics}, 50(10):102107, 2009.

\bibitem{saxena2021distributed}
Shivam Saxena, Hany E.~Z. Farag, and Nader El-Taweel.
\newblock A distributed communication framework for smart grid control applications based on data distribution service.
\newblock {\em Electric Power Systems Research}, 201:107547, 2021.

\bibitem{seiwerth2025extending}
Corinna Seiwerth, Nurbek Halikulov, and Reinhard German.
\newblock Extending a data-centric distributed simulation framework for the energy domain.
\newblock \url{https://doi.org/10.5281/zenodo.15064826}, 2025.
\newblock Accessed: 2026-03-28.

\bibitem{anggara2024case}
Taufik~Rendi Anggara.
\newblock A case study of implementation strategy for performance optimization in distributed cluster system.
\newblock {\em Journal of Computer Science and Technology}, 2024.

\bibitem{rak2021own}
Tomasz Rak and Łukasz Schiffer.
\newblock Own hpc cluster based on virtual operating system.
\newblock In P.K. Mallick, A.K. Bhoi, G.~Marques, and Victor Hugo~C. de~Albuquerque, editors, {\em Cognitive Informatics and Soft Computing}, page Unknown. Springer, 2021.
\newblock Chapter.

\bibitem{tong2018efficient}
Qiuhui Tong, Xiu Li, and Bo~Yuan.
\newblock Efficient distributed clustering using boundary information.
\newblock {\em Neurocomputing}, 275:2476--2485, 2018.

\bibitem{ronkko2024premises}
Jami Rönkkö, Olli Ahonen, Ville Bergholm, Alessio Calzona, et~al.
\newblock On-premises superconducting quantum computer for education and research.
\newblock {\em EPJ Quantum Technology}, 11(1):Article 243, 2024.

\bibitem{caleffi2024distributed}
Marcello Caleffi, Michele Amoretti, Davide Ferrari, Daniele Cuomo, Jessica Illiano, and Antonio Manzalini.
\newblock Distributed quantum computing: a survey.
\newblock {\em Computer Networks}, 242:110221, 2024.

\bibitem{barral2024review}
David Barral et~al.
\newblock Review of distributed quantum computing: From single qpu to high performance quantum computing.
\newblock {\em Computer Science Review}, 49:100747, 2024.

\bibitem{diadamod2021distributed}
Stephen DiAdamo, Marco Ghibaudi, and James Cruise.
\newblock Distributed quantum computing and network control for accelerated vqe.
\newblock {\em IEEE Transactions on Quantum Engineering}, 2:1--21, 2021.

\bibitem{benjamin2006brokered}
Simon~C. Benjamin, Daniel~E. Browne, Joseph Fitzsimons, and John J.~L. Morton.
\newblock Brokered graph-state quantum computing.
\newblock {\em New Journal of Physics}, 8(8):141, 2006.

\bibitem{mpi40_standard}
{Message Passing Interface Forum}.
\newblock {MPI}: A message-passing interface standard.
\newblock Technical report, Message Passing Interface Forum, June 2021.

\bibitem{haner2021distributed_qmpi}
Thomas H{\"a}ner, Damian~S. Steiger, Torsten Hoefler, and Matthias Troyer.
\newblock Distributed quantum computing with {QMPI}.
\newblock In {\em 2021 IEEE International Conference on Quantum Computing and Engineering (QCE)}, pages 1--10. IEEE, 2021.

\bibitem{shi2023reference_qmpi}
Yuxiang Shi, Tommy Nguyen, Samuel Stein, Tim Stavenger, Marvin Warner, and Martin Roetteler.
\newblock A reference implementation for a quantum message passing interface.
\newblock In {\em Proceedings of the SC '23 Workshops of the International Conference on High Performance Computing, Network, Storage, and Analysis}, pages 170--176. Association for Computing Machinery, 2023.

\bibitem{cardama2025netqmpi}
F.~Javier C{\'a}rdama and Tom{\'a}s~F. Pe{\~n}a.
\newblock Netqmpi: An {MPI}-inspired software for programming distributed quantum applications over {NetQASM} sdk.
\newblock In {\em Proceedings of the 2025 IEEE International Conference on Cluster Computing Workshops (CLUSTER Workshops)}, pages 1--12. IEEE, 2025.

\bibitem{javadiabhari2024qiskit}
Ali Javadi-Abhari, Matthew Treinish, Kevin Krsulich, Paul~D. Nation, Lev~S. Bishop, Sarah Sheldon, Christopher~J. Wood, Jake Lishman, Julien Gacon, Piotr Czarnik, et~al.
\newblock Quantum computing with {Qiskit}.
\newblock {\em arXiv preprint arXiv:2405.08810}, 2024.

\bibitem{kim2023cudaq}
Jin-Sung Kim, Alex McCaskey, Bettina Heim, Manish Modani, Sam Stanwyck, and Timothy Costa.
\newblock {CUDA Quantum}: The platform for integrated quantum-classical computing.
\newblock In {\em Proceedings of the 2023 60th ACM/IEEE Design Automation Conference (DAC)}, pages 1--4. IEEE, 2023.

\bibitem{mccaskey2021extending_cpp}
Alexander~J. McCaskey, Per Lynggaard, Eugene~F. Dumitrescu, Raphael~C. Pooser, Andrew Smith, et~al.
\newblock Extending {C++} for heterogeneous quantum-classical computing.
\newblock {\em ACM Transactions on Quantum Computing}, 2(2):1--30, 2021.

\bibitem{cross2022openqasm3}
Andrew~W. Cross, Lev~S. Bishop, Sarah Sheldon, Paul~D. Nation, and Jay~M. Gambetta.
\newblock {OpenQASM 3}: A broader and deeper quantum assembly language.
\newblock {\em ACM Transactions on Quantum Computing}, 3(3):1--50, 2022.

\bibitem{qir2024spec}
{The QIR Alliance}.
\newblock {Quantum Intermediate Representation (QIR)} specification, 2024.
\newblock [Online].

\bibitem{nikolopoulos_efficient_2018}
Konstantinos~F. Nikolaou.
\newblock {\em Efficient Private Information Retrieval}.
\newblock PhD thesis, City University of New York, 2018.
\newblock Available at \url{https://academicworks.cuny.edu/gc_etds/4212}.

\bibitem{hu_demystifying_2025}
Zhiyi Hu, Siyuan Shen, Tommaso Bonato, and {others}.
\newblock Demystifying nccl: An in-depth analysis of gpu communication protocols and algorithms.
\newblock {\em arXiv preprint arXiv:2507.04786}, 2025.

\bibitem{galindo2002information}
A.~Galindo and M.~A. Martin-Delgado.
\newblock Information and computation: Classical and quantum aspects.
\newblock {\em Reviews of Modern Physics}, 74(2):347--423, apr 2002.
\newblock Also available as arXiv:quant-ph/0112105.

\bibitem{aktar2024scalable}
Shamminuj Aktar, Andreas Bärtschi, Abdel-Hameed~A. Badawy, and Stephan Eidenbenz.
\newblock Scalable experimental bounds for entangled quantum state fidelities.
\newblock {\em ACM Transactions on Quantum Computing}, 5(4):27:1--27:21, 2024.
\newblock Also available as arXiv:2210.03048.

\bibitem{bruss1999entanglement}
Dagmar Bruss.
\newblock Entanglement splitting of pure bipartite quantum states.
\newblock {\em Physical Review A}, 60(4):4344, oct 1999.
\newblock arXiv:quant-ph/9902023.

\end{thebibliography}

\end{document}